\documentclass[fleqn]{sn-jnl}

\usepackage[fontsize=11pt]{scrextend}
\usepackage{caption}
\usepackage[square,comma,numbers,sort&compress]{natbib}

\renewcommand{\abstractheadfont}{\normalfont\normalsize\bfseries\titraggedcenter}

\usepackage{xcolor,color}
\usepackage{bm,amsmath,amssymb,dsfont}
\usepackage{graphicx}
\usepackage{afterpage}
\usepackage{enumerate}
\usepackage{manyfoot}

\long\def\comment#1{ }

\newcommand{\snn}{s_{\scriptscriptstyle N\!N}}
\newcommand{\eqn}[1]{Eq.~\eqref{#1}}
\newcommand{\beq}{\begin{equation}}
\newcommand{\eeq}{\end{equation}}
\newcommand{\bal}{\begin{align}}
\newcommand{\eal}{\end{align}}

\newcommand{\nn}{\nonumber\\}
\newcommand{\rmd}{{\rm d}}
\newcommand{\dif}{{\rm d}}

\newcommand{\bk}{\bm{k}}

\newcommand{\xbj}{x_{_{\rm Bj}}}
\newcommand{\mcal}{\mathcal}
\newcommand{\bK}{\bm{K}}
\newcommand{\bP}{\bm{P}}

\begin{document}

\title[]{\bf Probing gluon saturation via diffractive jets in ultra-peripheral nucleus-nucleus collisions}

\author*[1]{\fnm{E.} \sur{Iancu}}\email{edmond.iancu@ipht.fr}
\author[2]{\fnm{A.H.} \sur{Mueller}}\email{ahm4@columbia.edu}
\author[3]{\fnm{D.N.} \sur{Triantafyllopoulos}}\email{trianta@ectstar.eu}
\author[4]{\fnm{S.Y.} \sur{Wei}}\email{shuyi@sdu.edu.cn}

\affil*[1]{\it Institut de physique th\'{e}orique, Universit\'{e} Paris-Saclay, CNRS, CEA, F-91191, Gif-sur-Yvette, France}
\affil[2]{\it Department of Physics, Columbia University, New York, NY 10027, USA}
\affil[3]{\it European Centre for Theoretical Studies in Nuclear Physics and Related Areas (ECT*)\\
and Fondazione Bruno Kessler, Strada delle Tabarelle 286, 38123 Villazzano (TN), Italy}
\affil[4]{\it Key Laboratory of Particle Physics and Particle Irradiation (MOE), Institute of frontier and interdisciplinary science, Shandong University, Qingdao, Shandong 266237, China}

\abstract{\normalsize We argue that semi-inclusive photo-production of a pair of hard jets via coherent diffraction in
nucleus-nucleus ultra-peripheral collisions at high energy
is a golden channel to study gluon saturation. The dominant contribution is the diffractive production 
of three jets in an asymmetric configuration. Two of the jets are hard and propagate at  nearly
central pseudo-rapidities. 
The third jet is semi-hard, with transverse momentum comparable to the nuclear saturation momentum, 
and is well separated in pseudo-rapidity from the hard dijets. 
The emission of the semi-hard jet allows for strong scattering, thus avoiding the ``higher-twist'' suppression of
the exclusive dijet production due to colour transparency. We compute the trijet cross-section using
the diffractive TMD factorisation which emerges from the CGC effective theory at high energy. 
The cross-section is controlled by gluon
saturation, which leaves its imprints on the structure of the final state, notably on the rapidity distribution.}

\maketitle

\section{Introduction}
\label{sect:intro}

For sufficiently high energies, hadronic interactions are expected to probe a dense and weakly coupled form 
of gluonic matter, known as the Colour Glass Condensate (CGC), which is made with ``small--$x$ gluons''
(i.e. gluons which carry small fractions $x\ll 1$ of the hadron longitudinal momentum) and whose
main characteristic is {\it gluon saturation} --- the fact that gluon occupation numbers are limited 
to values of order $1/\alpha_s$ by the gluon mutual interactions 
\cite{Iancu:2002xk,Iancu:2003xm,Gelis:2010nm,Kovchegov:2012mbw}. 
The most direct way to experimentally study this form of matter would be through strong scattering in the vicinity of
the  ``black disk'' limit (the unitarity limit for the scattering amplitudes). In the CGC picture at weak coupling,
unitarity corrections are depicted as multiple scattering off strong colour fields representing the saturated gluons.
These corrections are particularly important if the scattering probes sufficiently large transverse separations
in the hadronic target, $r\sim 1/Q_s$, with $Q_s$ 
the target saturation momentum  --- the typical transverse momentum of the saturated gluons. Conversely, a
small hadronic projectile with transverse size $r \ll 1/Q_s$ scatters only weakly,
by colour transparency, and probes the tail of the gluon distribution at large transverse momenta $k_\perp\gg Q_s$.
The saturation momentum $Q_s$ rises rapidly with $1/x$, due to the abundant production of soft gluons via bremsstrahlung,
and also with the  nuclear mass number $A$ for a nuclear target with $A\gg 1$: 
roughly,  $Q_s^2(x, A)\sim A^\delta/x^\lambda$ with $\lambda\sim 0.2$ and $\delta\sim 1/3$.
For $A\simeq 200$  (lead or gold nuclei) and $x\lesssim 10^{-2}$, one expects a semi-hard value $Q_s^2(x,A)\gtrsim 2$~GeV$^2$,
for which perturbative QCD should be (at least marginally) applicable.

The simplest hadronic probe that one can think of, is the quark-antiquark ($q\bar q$), colour-dipole, fluctuation of 
the exchanged photon in photon-mediated processes, like electron-nucleus deep inelastic scattering (DIS) and
nucleus-nucleus ($AA$)  ultra-peripheral collisions (UPCs). By properly selecting the structure and the kinematics
of the final state, one can adjust the spatio-temporal resolution of the colour dipole. In DIS, the transverse size
of the $q\bar q$ pair is controlled by the photon virtuality, $r^2\sim 1/Q^2$, whereas  the Bjorken variable $\xbj=Q^2/(2P\cdot q)$
fixes the gluon longitudinal momentum fraction: $x=\xbj$. In UPCs, the photon is quasi-real
($Q^2\approx 0$), yet the dipole size $r$ can be forced to be small by measuring a pair of hard jets in the final state
--- those initiated by the original $q\bar q$ pair.  These jets propagate nearly back-to-back in the transverse plane
and their transverse relative momentum $P_\perp$ fixes $r^2\sim 1/P_\perp^2$.

Clearly, such processes are sensitive to gluon saturation whenever the controlling scales $Q^2$ and/or 
$P_\perp^2$ are {\it semi-hard}, of the order of the target saturation momentum $Q_s^2(x, A)$.
Yet, this is not the most interesting situation in practice. As already mentioned, the scale $Q_s^2$ is only
marginally perturbative, hence larger dipoles with size $r\gtrsim 1/Q_s$ (that would contribute to inclusive DIS
when $Q^2\sim Q_s^2$) are not under control in perturbation theory. Also, semi-hard jets with $P_\perp\sim Q_s$
cannot be reconstructed as genuine jets in the calorimeters of the Large Hadron Collider (LHC). So, it is important to also envisage the possibility
to study saturation via harder processes, with $Q^2,\, P_\perp^2\gg Q_s^2$.

In a couple of recent papers \cite{Iancu:2021rup,Iancu:2022lcw},
 we have argued that the semi-inclusive photo-production of a pair of hard jets via coherent diffraction is a promising channel
in that sense. By ``coherent diffraction'' we mean elastic processes in which the target nucleus is not broken by
the collision. Elastic scattering is indeed well suited for a study of gluon saturation since particularly sensitive to  unitarity
corrections. It is generally revealed by the presence of large rapidity gaps in the final state.
By ``semi-inclusive'' we refer to events where the hard $q\bar q$ dijets, with relative momentum $P_\perp\gg Q_s$, are accompanied by (at least) one semi-hard jet --- a gluon with transverse momentum $K_\perp\sim Q_s$ --- whose emission 
by either the quark or the antiquark plays an essential role in the economy of the process: it re-distributes colour over a large transverse separation $R\sim 1/Q_s$, thus allowing for strong scattering.
Without this third jet, the cross-section for the {\it exclusive} production of the hard dijets would be strongly suppressed, 
by a factor $Q_s^2/P_\perp^2$, due to colour transparency (see Appendix~\ref{sec:app} below for an explicit calculation).

The analysis in  \cite{Iancu:2021rup,Iancu:2022lcw} also led to  interesting conceptual clarifications: 
the cross-section for diffractive ``(2+1)-jet production" (two hard jets accompanied by a semi-hard one) 
admits {\it transverse-momentum dependent (TMD)
factorisation}: it can be written as the product between a ``hard factor'' describing the hard dijet production and a ``semi-hard'' factor,
representing the unintegrated gluon distribution of the ``Pomeron'' (the colourless exchange between
the nuclear target and the three jets produced in the final state), also known as (a.k.a.) the gluon diffractive TMD.
Whereas the emergence of collinear factorisation for diffraction from the dipole picture was anticipated in the early works  \cite{Wusthoff:1997fz,GolecBiernat:1999qd,Hebecker:1997gp,Buchmuller:1998jv,Hautmann:1998xn,Hautmann:1999ui,
Hautmann:2000pw}, the TMD factorisation exhibited in   \cite{Iancu:2021rup,Iancu:2022lcw} is nevertheless remarkable in several respects. First, it holds at the ``unintegrated'' level, that is, for a fixed value of the transverse momentum $K_\perp$ of the gluon jet (see also \cite{Hatta:2022lzj,Beuf:2022kyp} for recent developments).
Second, the gluon diffractive TMD is shown to be controlled by the ``semi-hard'' physics of gluon saturation ($K_\perp\sim Q_s$), hence it 
can be computed from first principles. This is particularly important for applications to heavy nuclei, 
for which the diffractive parton distributions are 
only poorly known and previous studies based on collinear factorisation 
had to resort on models \cite{Frankfurt:2011cs,Guzey:2016tek}.

Our original analysis focused on the DIS processes to be studied at the Electron-Ion Collider (EIC)
\cite{Accardi:2012qut,Aschenauer:2017jsk,AbdulKhalek:2022hcn}. In the present paper we extend this analysis 
to diffractive (2+1)-jet production in  $AA$ UPCs \cite{Bertulani:1987tz,Bertulani:2005ru,Baltz:2007kq},
with emphasis on the kinematical conditions at the LHC. While most of the previous,
theoretical and experimental, efforts have been devoted to vector meson photo-production   
(see e.g. \cite{Contreras:2015dqa,Klein:2019qfb} and references therein), there are also recent measurements 
of diffractive dijets in Pb+Pb UPCs at the LHC  \cite{ATLAS:2017kwa,ATLAS:2022cbd,CMS:2020ekd,CMS:2022lbi}. 

A priori, the LHC looks better suited than the EIC for a study of gluon saturation,
due to the much higher available energies. In UPCs though, this advantage is somewhat diminished by the limitation
on the energy of the exchanged photon (which is only a small fraction of that of its parent nucleus) and by the fact that
the measured jets are quite hard, $P_\perp \ge 20$~GeV, and hence cannot access
very small values of $x$. (The relevant $x$--variable in this diffractive context is the longitudinal momentum fraction 
$x_{\mathbb{P}}$ lost by the nuclear target and taken by the Pomeron.) The smallest values
of  $x_{\mathbb{P}}$ explored by the dijet measurements in \cite{ATLAS:2017kwa,ATLAS:2022cbd,CMS:2020ekd,CMS:2022lbi},
in the ballpark of $x_{\mathbb{P}}=0.01$, are only marginally favourable for saturation --- for instance, they do not allow one
to test the high-energy evolution in QCD, as described by the BK/JIMWLK equations
\cite{Balitsky:1995ub,JalilianMarian:1997jx,JalilianMarian:1997gr,Kovner:2000pt,Iancu:2000hn,Iancu:2001ad,Ferreiro:2001qy,Kovchegov:1999yj}. That said, for $x_{\mathbb{P}}=0.01$ and a large target nucleus like Pb ($A=208$), 
one may still expect saturation effects in the sense of the  semi-classical McLerran-Venugopalan (MV) model  \cite{McLerran:1993ni,McLerran:1994vd}. This is the scenario that we shall privilege in this paper.

The highly asymmetric structure of the (2+1)--jet final state introduces both experimental and conceptual challenges.
A semi-hard jet with transverse momentum $K_\perp$ of the order of $Q_s$  is rather hard to measure --- e.g., it is too soft
to be reconstructed in a calorimeter at the LHC. As a matter of fact, no such a jet was reported by the recent analysis by CMS \cite{CMS:2020ekd,CMS:2022lbi}.  Yet, as we shall argue in this paper, the actual 
observation of this jet (say, via its hadronic descendants) would be highly beneficial. First, it would allow one to 
unambiguously distinguish the nucleus which emitted the photon from that which acted as a target. Second, it would
permit to measure the diffractive gap, which (in our calculation at least) is the rapidity gap between the gluon jet and
the nuclear target. So it is  important to understand where to look for this third jet in the final event. 
Our analysis demonstrates that the hard dijets are predominantly produced at relatively central pseudo-rapidities
($|\eta| \lesssim 1$) and that they are separated from the semi-hard jet by a rather large interval 
$\Delta\eta_{\rm jet}\sim \ln(P_\perp/Q_s)$ (2 to 3 units of pseudo-rapidity). Hence, the third jet is likely  
to propagate at large pseudo-rapidities, which may explain why it was not detected by CMS.
Since controlled by the target saturation momentum $Q_s$, this large rapidity separation $\Delta\eta_{\rm jet}$
brings direct evidence for gluon saturation.

But the most important evidence in favour of saturation is the very fact that the coherent (diffractive) production of a pair of hard
dijets has a large cross-section, which is of leading-twist order at large $P_\perp$ and proportional to $A$
(the mass number of the target nucleus), and hence of the same order as
the cross-section for inclusive dijet photo-production via inelastic processes. By contrast, the cross-section for exclusive
dijets is of higher twist order (see Appendix~\ref{sec:app}); for the experimental conditions at the LHC, it should be down
by 2 to 3 orders of magnitude compared to the semi-inclusive cross-section corresponding to the (2+1)-jet channel.

\section{Diffractive dijets in UPCs}
\label{sec:ddUPC}

We shall study the diffractive production of three jets via $\gamma A$ interactions in coherent ultra-peripheral nucleus-nucleus collisions (UPCs). Ultra-peripheral means that the impact parameter of the collision is larger than the sum $R_A+R_B$ of the two nuclear radii, while by ``coherent'' we mean
that both nuclei survive in the final state. Although usually the nuclei are identical, we will 
use different labels to distinguish the nucleus which propagates in the $+z$ direction (``right-mover''), denoted as $B$, from the left-mover, denoted as $A$.
The nuclei are ultra-relativistic, so that the center-of-mass energy squared for a nucleon-nucleon collision reads $\snn = 2 P_B^+ P_A^- = 4 E_{N}^2$, where $E_N \gg M_N$ is the energy of a nucleon with mass $M_N$, while $P_B^+ = \sqrt{2} E_N$ is the large longitudinal momenta of a nucleon in nucleus $B$ and $P_A^- = \sqrt{2} E_N$ the respective momentum for a nucleon in nucleus $A$. 
For quantitative estimates we shall take $\sqrt{\snn} = 5$ TeV and thus $E_N = 2.5$ TeV, like for Pb+Pb collisions at the LHC. 

\begin{figure}
	\begin{center}
		\includegraphics[width=0.56\textwidth]{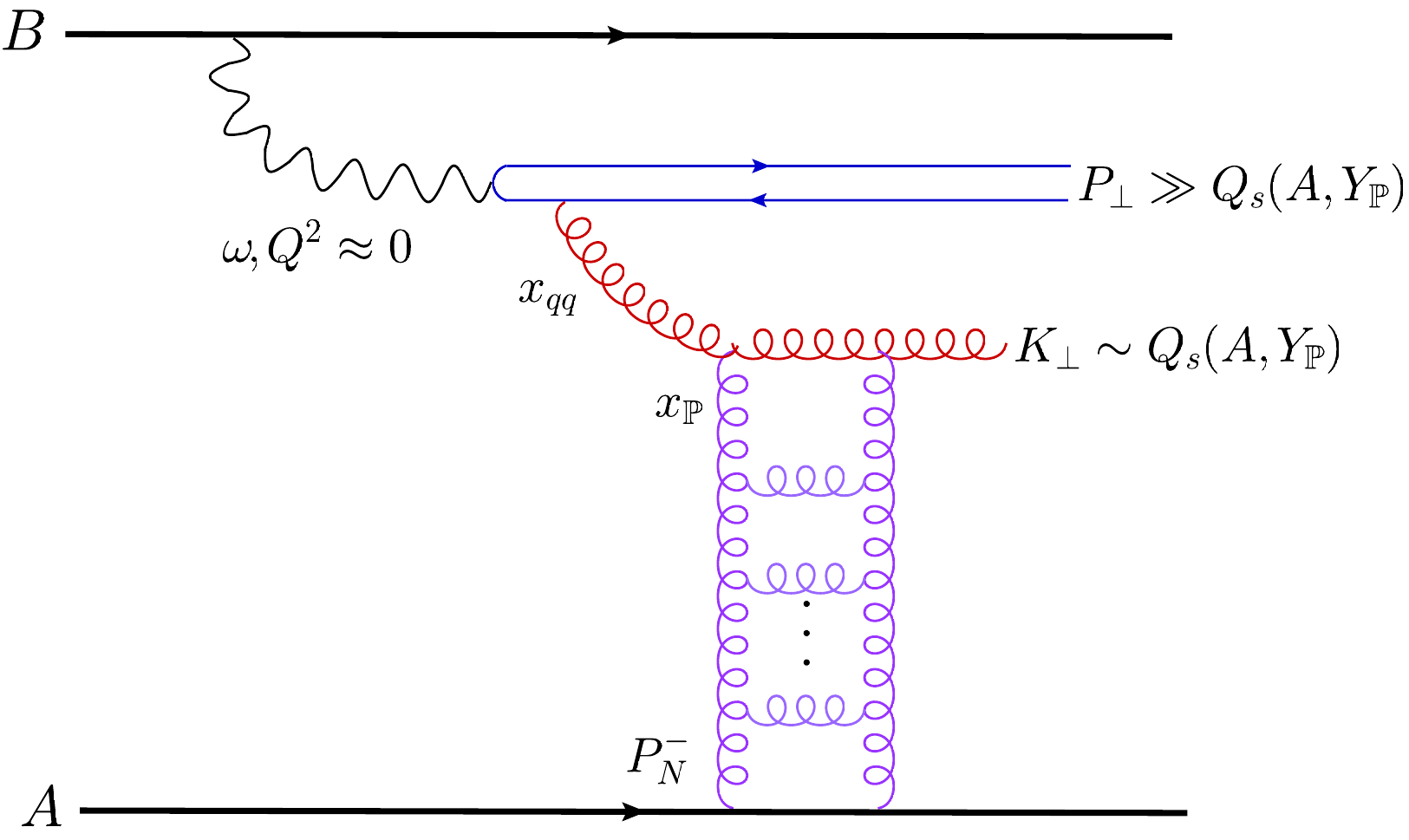}
	\end{center}
	\caption{Diffractive photo-production of 2+1 jets (a hard quark-antiquark pair with large relative transverse momentum $P_\perp$
	and a semi-hard gluon with semi-hard $K_\perp\sim Q_s(A,Y_{\mathbb{P}})$) in the coherent ultraperipheral collision.}
\label{fig:UPC}
\end{figure}

One of the nuclei acts as a source for the quasi-real photon $\gamma$ and the other one is the target off which the photon scatters.
 For definiteness, we will work out the case in which the photon is emitted from nucleus $B$ and hence it is a right-mover too. The other case (photon emitted from nucleus $A$) can be trivially deduced by a symmetry operation (see below) and the final cross section is obtained by adding the two possibilities. The photon has a very small space-like virtuality and zero transverse momentum, thus we can write its 4-momentum as $q^{\mu}= (q^+\simeq \sqrt{2}\omega,q^-=-Q^2/2q^+, \bm{0}_{\perp})$, with $|q^-|\ll q^+$. The process which we are interested in is the following: first the photon decays to a quark-antiquark pair and then a gluon is emitted either from the quark or the antiquark (see Fig.~\ref{fig:UPC} for a pictorial representation). 
Then the three partons elastically scatter off the nucleus $A$ and emerge in the final state with
4-momenta $k_i^{\mu}=(k_i^+,k_i^-,\bk_{i\perp})$, where $k_i^{-} = k_{i\perp}^2/2k_i^+$ and the labels $i=1,2,3$ refer to the quark, the antiquark and the gluon respectively. It is useful to define the longitudinal fractions $\vartheta_i = k_i^+/q^+$ with respect to (w.r.t.) the photon and similarly  $x_i = k_i^-/P_N^-$ w.r.t.~the struck nucleon. We clearly have $\vartheta_1+\vartheta_2+\vartheta_3 =1$, whereas the sum
\begin{align}
	\label{xpom}
	x_{\mathbb{P}} = x_1+ x_2 + x_3
\end{align}
is the fraction of $P_N^-$ transferred from the target nucleus $A$ to the produced jets via the ``Pomeron'' (cf.~the detailed discussion at the end of the current section). In diffraction there shouldn't be any net color flow between the projectile partons and the nuclear target, or in other words the Pomeron must be colorless.  Then one should observe a pseudo-rapidity gap, i.e.~an angular region between the target nucleus and the produced jets, which is void of particles. The value $\Delta \eta_{\rm gap}$ of this diffractive gap is close (but not exactly equal) to the rapidity interval $Y_{\mathbb{P}} = \ln 1/x_{\mathbb{P}}$ relevant for the high energy evolution of the target nucleus $A$. We are interested in the case that the Pomeron longitudinal momentum fraction be small enough, say $x_{\mathbb{P}} \lesssim 0.01$, in order to probe gluon saturation, that is, to have a semi-hard value for the saturation momentum $Q_s(Y_{\mathbb{P}})$ of the target nucleus $A$.  The condition $x_{\mathbb{P}} \ll 1$ is also needed to allow for coherent scattering, that is,  to ensure
 that nucleus $A$ emerges unbroken from the collision.
 
As explained in the Introduction, we are interested in asymmetric 3-jet configurations, where two of the jets,
here chosen as  the $q\bar q$ pair, are much harder than the third (gluon) jet.
(The case where one of the two hard jets is the gluon will be addressed in a further study.) Specifically, the transverse momenta  
$k_{1\perp}$ and $k_{2\perp}$ of the quark and the antiquark are assumed to be large compared to the saturation momentum 
$Q_s(Y_{\mathbb{P}})$ and their longitudinal momentum
fractions $\vartheta_1$ and $\vartheta_2$ take generic values of the order of one half. Then, as explained in  \cite{Iancu:2021rup,Iancu:2022lcw},
 the gluon jet is dynamically selected to be semi-hard, i.e.~$k_{3\perp} \sim Q_s(Y_{\mathbb{P}})$, and soft, namely $\vartheta_3 \lesssim k_{3\perp}^2/k_{1\perp}^2$. The first condition on the transverse momentum makes sure that the gluon's distance $R \sim 1 /k_{3\perp}$ from the $q\bar{q}$ pair is large enough to allow for strong scattering, while the upper limit on the fraction $\vartheta_3$ guarantees that the gluon formation time does not exceed 
the $q\bar{q}$ pair lifetime. If any of these two constraints is not satisfied, the cross section of interest is strongly suppressed.

For the coherent process under consideration,
the total transverse momentum $\Delta_{\perp}$ transferred from the target to the jets is very small, $\Delta_{\perp}\sim 1/R_A \simeq 30$ MeV
(we used $R_A \simeq 6$ fm for a Pb nucleus). This  is much smaller than any of  
the transverse momenta of the three jets. Accordingly, the momentum imbalance between the 
hard jets is determined  by the gluon, $|\bk_{1\perp}+\bk_{2\perp}| \simeq k_{3\perp}$, and is much smaller then their individual momenta 
$k_{1\perp}$ and $k_{2\perp}$.  This makes it clear that the two hard jets propagate nearly  back-to-back in the transverse plane.
It is then convenient to introduce the relative ($\bP$) and total ($\bK$)  momenta of the hard dijet according to
\begin{align}
	\label{pandk}
	\bP \equiv \frac{\vartheta_2 \bk_1 - 
	\vartheta_1 \bk_2}
	{\vartheta_1+\vartheta_2},
	\qquad
	\bK \equiv \bk_1 + \bk_2
\end{align}
and the kinematic regime of interest can be summarised as $k_{1\perp} \simeq  k_{2\perp} \simeq P_{\perp} \gg K_{\perp} \simeq k_{3\perp} \simeq Q_s(Y_{\mathbb{P}})$ and $\vartheta_1 \sim \vartheta_2 \simeq 1-\vartheta_1 \gg \vartheta_3$.

It is standard to use (pseudo-)rapidities to characterise the jets longitudinal momenta. 
For massless particles there is no difference between  pseudo-rapidities and rapidities, so we will use the common notation $\eta$ for both. For an on-shell particle with 4-momentum $k^{\mu} $  we have 
 \begin{align}
 	\label{rapid}
 	\eta = -\ln \tan \frac{\theta}{2} =
 	\frac{1}{2}\,\ln \frac{k^+}{k^-} =
 	 \ln \frac{\sqrt{2} k^+}{k_{\perp}}
 	 =- \ln \frac{\sqrt{2} k^-}{k_{\perp}},
 \end{align}
where $\theta$ is the propagation angle w.r.t.~the $z$-axis. We shall say that a particle moves in the forward direction when its rapidity is positive, i.e.~when it propagates in the same hemisphere as the nucleus $B$ and the photon emitted by the latter.

We are now in a position to write down the general expression for the cross section for 2+1 jet production in $AB$ UPCs:  
\begin{align}
	\label{sigmaAB}
	\frac{\dif \sigma_{2+1}^{AB\to q\bar{q}gAB}}
	{\dif\eta_1 \dif \eta_2\dif^2\bP\dif^2\bK \dif Y_{\mathbb{P}}} = 
	\int_0^\infty
	\dif \omega 
	\bigg[ \frac{\dif N_B}{\dif \omega}
	\frac{\dif \sigma_{\rm D}^{\gamma A \to q\bar{q}gA}}
	{\dif\eta_1 \dif \eta_2\dif^2\bP\dif^2\bK \dif Y_{\mathbb{P}}}
	+(A \leftrightarrow B)
	\bigg],
\end{align}
where the first (second) term in the square brackets refers
to the case in which the photon is emitted by nucleus $B$ (respectively, $A$).  As already mentioned, we will 
focus on the first case --- photon emitted by $B$ and which scatters with $A$ ---,
 for which we shall use the label $BA\to \gamma A$. The  contribution of the other case ($AB\to \gamma B$) 
can then be obtained by changing the signs of the pseudo-rapidities $\eta_1$ and $\eta_2$. 
Each of these terms is the product of two factors  which 
describe the two stages of the process: the photon emission and the photon-nucleus collision.

\begin{figure}
	\begin{center}
		\includegraphics[width=0.46\textwidth]{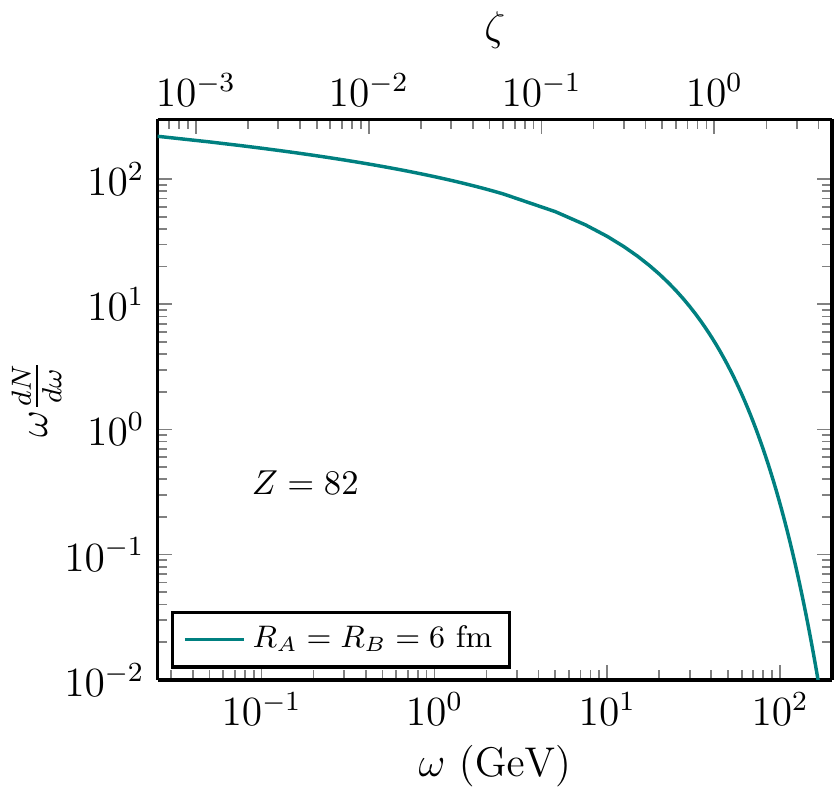}
		\hspace*{0.06\textwidth}
		\includegraphics[width=0.46\textwidth]{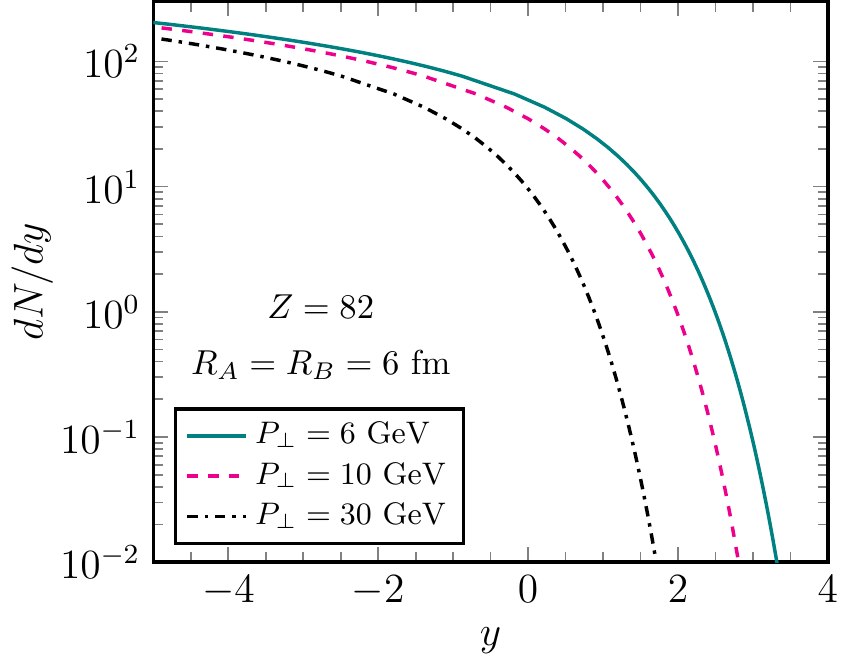}
	\end{center}
	\caption{Left panel: Double logarithmic plot of the photon spectrum produced by an ultra-relativistic nucleus with charge $Z=82$, radius $R_A=6$ fm and Lorentz factor $\gamma_L=2500$, passing a same target nucleus at an impact parameter $b \geq 2 R_A$, either as a function of the photon energy $\omega$ or as a function of the dimensionless parameter $\zeta$ (cf.~Eq.~\eqref{zeta}).  Right panel: Logarithmic plot of the same spectrum as a function of the common rapidity $y=\eta_1=\eta_2=\ln (\omega/P_{\perp})$ of a symmetric dijet pair with transverse momenta $k_{1\perp} = k_{2\perp} = P_{\perp}$.}
\label{fig:flux}
\end{figure}

The quantity $\dif N_B/\dif \omega$ is obtained by integrating the photon flux generated by nucleus $B$ over impact parameters $b \geq R_A + R_B$. A computation in classical electrodynamics gives \cite{Jackson:1998nia,Bertulani:1987tz,Bertulani:2005ru,Baltz:2007kq}
\begin{align}
	\label{flux}
	\frac{\dif N_B}{\dif \omega} = 
	\frac{2 Z_B^2 \alpha_{\rm em}}{\pi \omega}
	\bigg\{
	\zeta K_0(\zeta) K_1(\zeta) - \frac{\zeta^2}{2}
	\big[ K_1^2(\zeta) - K_2^2(\zeta) \big] 
	\bigg\},
\end{align}
with $\alpha_{\rm em}$ the fine structure constant and where we have defined the dimensionless parameter
\begin{align}
	\label{zeta}
	\zeta=\frac{\omega (R_A +R_B)}{\gamma_L} = 2 x_{\gamma} M_N R_A.
\end{align}
Here $\gamma_L = E_N/M_N$ is the nucleon Lorentz boost factor, whereas in writing the second equality we assumed identical nuclei. We also introduced the fraction $x_{\gamma} = q^+/P_B^+ = \omega/E_N$ of the longitudinal momentum of a nucleon (from nucleus $B$) that is carried by the photon. We show the integrated photon flux as a function of both $\omega$ and $\zeta$ in the left panel of Fig.~\ref{fig:flux}. Since the Bessel functions vanish exponentially for $\zeta \gtrsim 1$, it becomes clear that the photon flux is substantial only for energy fractions up to 
\begin{align}
	\label{xmax}
	x_{\gamma}^{*} \equiv \frac{1}{2 M_N R_A}.
\end{align}
With $M_N = 1$ GeV and $R_A=6$ fm, one finds $x_{\gamma}^{*} \simeq 0.016$, which in turn implies a critical value $\omega^{*} = x_{\gamma}^{*} E_N \simeq$ 40 GeV for the photon energy. We would like to stress that higher photon energies are not kinematically forbidden and it would be very welcome if experiments could trigger on such rare events. The fact that $\omega^{*}$ is much smaller than the nucleon energy $E_N$ means that, although the photon is a right mover, the jets will not be constrained to move in the forward direction as we will discuss in a while.  

The integration over $\omega$ in Eq.~\eqref{sigmaAB} can be trivially performed by using the conservation of the plus component of the
longitudinal momentum in the $\gamma A$ collision: $q^+=k_1^++k_2^++k_3^+ \simeq k_1^++k_2^+$, where we have used the fact that the gluon 
is soft. Hence $\omega$ is determined by the final state kinematics of the hard pair according to (we recall that  $q^+\simeq \sqrt{2}\omega$)
\begin{align}
	\label{omega}
	\omega = \frac{1}{\sqrt{2}}\, (k_1^+ + k_2^+) = 
	\frac{1}{2}\,(k_{1\perp} e^{\eta_1} + k_{2\perp} e^{\eta_2})\simeq
	\frac{P_{\perp} }{2}\,(e^{\eta_1} + e^{\eta_2}).
\end{align}

The second ingredient in Eq.~\eqref{sigmaAB} is the cross section for 2+1 jet production in $\gamma A$ coherent diffraction. When $P_{\perp} \gg K_{\perp}$ and $\vartheta_1,\vartheta_2 \gg \vartheta_3$, this factorises between a ``hard'' and a ``semi-hard'' factor
\cite{Iancu:2021rup,Iancu:2022lcw}, as we now explain. 
By taking into account only the first term in Eq.~\eqref{sigmaAB} we can write
\begin{align}
	\label{sigmafac}
	\frac{\dif \sigma_{2+1}^{BA\to \gamma A}}
	{\dif\eta_1 \dif \eta_2\dif^2\bP\dif^2\bK \dif Y_{\mathbb{P}}} =
	\omega\frac{\dif N_B}{\dif \omega}\,
	h(\eta_1,\eta_2,P_{\perp}^2)\,
	\frac{\dif x G_{\mathbb{P}}^A(x,x_{\mathbb{P}},K_{\perp}^2)}{\dif^2\bK},
\end{align}
where the photon energy is fixed according to \eqn{omega}.

The hard factor $h$ describes the decay of a transversely polarised  quasi-real photon\footnote{The contribution of longitudinally polarized photons vanishes linearly with $Q^2$.} into a $q\bar{q}$ pair, as well as the coupling of the latter to the comparatively soft gluon; it reads
\begin{align}
	\label{hfactor}
	h(\eta_1,\eta_2,P_{\perp}^2) = 
	\alpha_{\rm em} \alpha_s
	\left(\sum {e_f^2} \right) \vartheta_1 \vartheta_2 (\vartheta_1^2+ \vartheta_2^2)\, \frac{1}{P_{\perp}^4},
\end{align}
where $\alpha_s$ is the strong coupling constant and $e_f$ is the fractional charge of the flavor $f$. It is understood that $\vartheta_2 \simeq 1- \vartheta_1$ and $\vartheta_1$ can be expressed in terms of pseudo-rapidities as
\begin{align}
	\label{theta1}
	\vartheta_1 = \frac{k_{1\perp} e^{\eta_1}}
	{k_{1\perp} e^{\eta_1} + k_{2\perp} e^{\eta_2}}
	\simeq
	\frac{e^{\eta_1}}{e^{\eta_1}+e^{\eta_2}}.
\end{align}
The factor $\vartheta_1 \vartheta_2$ in Eq.~\eqref{hfactor} makes clear that relatively symmetric jets with $\vartheta_1 \sim \vartheta_2 \sim 1/2$, 
hence with $\eta_1 \sim \eta_2$, are favored by this process.

The semi-hard factor in Eq.~\eqref{sigmafac} encodes the QCD dynamics of interest and represents the unintegrated gluon distribution (UGD) of the Pomeron, or the ``gluon diffractive transverse momentum distribution (TMD)'': it gives the probability to find a gluon with relative minus longitudinal momentum fraction $x$ and transverse momentum $K_{\perp}$ inside a Pomeron which carries a fraction $x_{\mathbb{P}}$ of the target  momentum (per nucleon). Implicit 
in this interpretation, there is an alternative physical picture in which the gluon is viewed as being part of the target nucleus wavefunction\footnote{That
would be the actual physical picture if the $\gamma A$ collisions was viewed in a different frame and gauge: the Bjorken frame where the photon has
zero longitudinal momentum and the target light-cone gauge.} --- more precisely, of the ``Pomeron''.
The Pomeron emits two gluons in a color singlet state: one in the $s$--channel (which appears in the final state) with  momentum fraction $x_3$ 
and transverse momentum $\bk_3=-\bm{K}$, and one in the $t$--channel, with momentum fraction $ x_{\mathbb{P}} - x_3$
and transverse momentum $\bm{K}$. The $t$--channel gluon is absorbed by the $q\bar{q}$ pair and provides the minus momentum fraction of the
hard dijet,  $x_{\mathbb{P}} -x_3 = x_1+x_2 \equiv x_{q\bar{q}}$, as well as its transverse momentum imbalance: $\bm{K}=\bk_1+\bk_2$.
The variable $x$ appearing in the gluon diffractive TMD is the splitting fraction of the $t$--channel gluon w.r.t. the Pomeron:
\begin{align}
	\label{x}
	x = \frac{x_{q\bar{q}}}{x_{\mathbb{P}}}.
\end{align}
While $x_{q\bar{q}}$ depends only on the kinematics of the hard dijet, namely 
\begin{align}
	\label{xqq}
	x_{q\bar{q}} = \frac{k_{1\perp} e^{-\eta_1}+k_{2\perp} e^{-\eta_2}}{2 E_N}
	\simeq \frac{P_{\perp}}{2 E_N}
	\left(e^{-\eta_1}+e^{-\eta_2} \right),
\end{align}
$x$ and $x_{\mathbb{P}}$ are also sensitive to the kinematics of the semi-hard gluon jet, more precisely 
\begin{align}
	\label{xpom}
	x_{\mathbb{P}} = x_{q\bar{q}}\, + \frac{K_{\perp} e^{-\eta_3}}{2 E_N}
\end{align}
and $x$ can be determined from the above equations. Notice that, out of the three longitudinal variables $\eta_3, \, x_{\mathbb{P}}$ and $x$,
only one is independent.

\section{The gluon distribution of the Pomeron}
\label{sec:Pomeron} 

As emphasised in the Introduction,  the Pomeron UGD at low $x_{\mathbb{P}}\ll 1$
 is determined by the physics of gluon saturation. It is related to a particular Fourier-Bessel transform of the amplitude $\mcal{T}_{g}(R,Y_{\mathbb{P}})$ for a gluon-gluon dipole of size $R$ to scatter off the nuclear target. More precisely one finds \cite{Iancu:2021rup,Iancu:2022lcw}
 \begin{align}
 	\label{xgp-gen}
 	\frac{\dif x G_{\mathbb{P}}^A(x,x_{\mathbb{P}},K_{\perp}^2)}{\dif^2\bK}
 	= \frac{S_{\perp}(N_c^2-1)}{4\pi^3}\,
	\frac{\big|\mcal{G}^A(x,x_{\mathbb{P}},K_{\perp})\big|^2}{2\pi (1-x)},
 \end{align}
 with $S_{\perp}\propto A^{2/3}$ the transverse area of the nucleus $A$ and $\mcal{G}^A$ a
dimensionless distribution  given by
 \begin{align}
 	\label{Gcal}
 	\mcal{G}^A(x,x_{\mathbb{P}},K_{\perp}) = 
 	\mcal{M}^2 \int_0^{\infty}
 	\dif R\, R\,
 	J_2(K_{\perp} R)
 	K_2(\mcal{M}R)
 	\mcal{T}_{g}(R,Y_{\mathbb{P}})
 	\qquad
 	\textrm{with}
 	\qquad
 	\mcal{M}^2 \equiv \frac{x}{1-x}\,K_{\perp}^2.
 \end{align}
 For a qualitative discussion, we shall use the MV model  \cite{McLerran:1993ni,McLerran:1994vd}, which 
 is a reasonable approximation for a large nucleus and not too high energies. In this model,
 the  gluon-gluon dipole amplitude is independent of $Y_{\mathbb{P}}$ and reads
\begin{align}
	\label{MV}
	\mcal{T}_g(R)=
	1 - \exp \left( - \frac{Q_{gA}^2 R^2}{4}\,\ln \frac{4}{R^2 \Lambda^2} \right),
\end{align}
where $\Lambda$ is the QCD scale and $Q_{gA}^2$ is the colour charge density of the valence quarks in nucleus $A$, as measured by the gluon-gluon dipole, hence it is proportional to $N_c A^{1/3}$.  
%(Notice that this is naturally enhanced by a factor $N_c/C_F=9/4$, when compared to the respective scale appearing in the amplitude for the scattering of a $q\bar{q}$ dipole.) 
We define the gluon saturation momentum $Q_{gs}$ via the condition that the exponent in Eq.~\eqref{MV} becomes equal to one when $R=2/Q_{gs}$, which gives 
\begin{align}
	\label{Qgs}
	Q_{gs}^2 = Q_{gA}^2
	\ln \frac{Q_{gs}^2}{\Lambda^2}.
\end{align}
The Pomeron UGD coming out from the MV model can be given the piecewise 
form \cite{Iancu:2021rup,Iancu:2022lcw}
\begin{align}
	\label{xgpu}
	\frac{\dif x G_{\mathbb{P}}^A(x,x_{\mathbb{P}},K_{\perp}^2)}{\dif^2\bK}
	\simeq
	\frac{S_{\perp}(N_c^2-1)}{4\pi^3}\,
	\frac{1-x}{2\pi}
	\begin{cases}
		1 
		&\quad \mathrm{for}  \quad K_{\perp} \ll \tilde{Q}_s(x)
		\\*[0.2cm]
		{\displaystyle \frac{\tilde{Q}_s^4(x)}{K_{\perp}^4}}
		&\quad \mathrm{for} \quad K_{\perp} \gg \tilde{Q}_s(x).
	\end{cases} 
\end{align}
The expression in the first line arises from the unitarity limit $\mcal{T}_g=1$, while that in the second line, in which we have neglected slowly varying functions of $x$ and $K_{\perp}^2$, comes from the weak scattering limit $\mcal{T}_g \sim R^2 Q_{gs}^2$. In the above, $\tilde{Q}_s^2(x) = (1-x) Q_{gs}^2$ is an
\emph{effective} saturation momentum which can be understood as follows: the $t$-channel gluon has a space-like virtuality $k_g^2 = K_{\perp}^2/(1-x)$, so that the typical size squared of the gluon-gluon dipole is $R^2 \sim 1/k_g^2 = (1-x)/K_{\perp}^2$. Then the condition that the scattering be strong for $R \gtrsim 1/Q_{gs}$, leads to $K_{\perp} \lesssim \tilde{Q}_s(x)$.

The rapid $1/K_{\perp}^4$ fall-off in the tail of the Pomeron UGD implies that the transverse momentum of the third jet in the typical events
is of the order of $\tilde{Q}_s(x)$. In turn, this has two significant phenomenological consequences for UPCs: (i) the gluon jet is too soft to be observed as a genuine jet in a calorimeter, albeit its hadronic descendants could be still measured in a hadron detector,
and (ii) the dijet imbalance is eventually controlled by mechanisms which have not been taken yet into account, namely the DGLAP evolution 
\cite{Gribov:1972ri,Altarelli:1977zs,Dokshitzer:1977sg} of the Pomeron gluon distribution \cite{Iancu:2022lcw} (see below)
and the final state radiation from the hard dijet \cite{Hatta:2021jcd}. Such effects involve emissions of partons with transverse momenta 
logarithmically distributed between $\tilde{Q}_s(x)$ and $P_{\perp}$, so that the final dijet imbalance becomes considerably larger 
than the semi-hard scale $\tilde{Q}_s(x)$.

Since the experimentally measured dijet imbalance $K_{\perp}$ is not under control within the present approach, nor
representative for the physics of saturation, it is
preferable to integrate over $K_{\perp}$ up to the hard scale $P_{\perp}$ and thus obtain a {\it semi-inclusive} cross-section for hard dijet production
\begin{align}
	\label{sigmaint}
	\frac{\dif \sigma_{2+1}^{BA\to \gamma A}}
	{\dif\eta_1 \dif \eta_2\dif^2\bP\dif Y_{\mathbb{P}}} =
	\omega\frac{\dif N_B}{\dif \omega}\,
	h(\eta_1,\eta_2,P_{\perp}^2)\,
	x G_{\mathbb{P}}^A(x,x_{\mathbb{P}},P_{\perp}^2),
\end{align}  
which involves the {\it integrated} gluon distribution of the Pomeron $x G_{\mathbb{P}}^A(x,x_{\mathbb{P}},P_{\perp}^2)$, a.k.a.  
the gluon diffractive parton distribution function (DPDF).
It is suggestive to present its expression in the MV model, cf. \eqn{xgpu}. Then the integral is  dominated by momenta $K_{\perp}\sim\tilde{Q}_s(x)$ and is almost independent of the upper limit  $P_{\perp}$ so long as $P_{\perp} \gg \tilde{Q}_s(x)$. The result can be written as \cite{Iancu:2021rup,Hatta:2022lzj,Iancu:2022lcw}
\begin{align}
	\label{xgp}
	x G_{\mathbb{P}}^A(x,x_{\mathbb{P}},P_{\perp}^2) &= 
	\int_0^{P_{\perp}}
	\dif^2\bK\,
	\frac{\dif x G_{\mathbb{P}}^A(x,x_{\mathbb{P}},K_{\perp}^2)}{\dif^2\bK}
	\nn
	&= \frac{S_{\perp}(N_c^2-1)}{4\pi^3}\,
	\kappa\big( x, P_{\perp}^2/\tilde{Q}_s^2(x) \big)
	(1-x) \tilde{Q}_s^2(x),
\end{align}
where $\kappa$ is a slowly varying function in all of its arguments.  In this semi-classical approximation, valid at small 
$x_{\mathbb{P}}$, the gluon distribution of the Pomeron is independent of $x_{\mathbb{P}}$.
Since proportional to $\tilde{Q}_s^2(x) = (1-x) Q_{gs}^2$, the right hand side (r.h.s.) of \eqref{xgp} 
scales like  $A\ln A$ (recall \eqn{Qgs}) and vanishes like $(1-x)^2$ when $x\to 1$.
Both scaling laws are sensitive to gluon saturation: in the absence of saturation, the single-scattering approximation in the second
line of \eqn{xgpu} would apply for all transverse momenta down to $K_\perp\sim \Lambda$ and as a consequence 
$x G_{\mathbb{P}}^A$ would scale like $A^{4/3}$ and would vanish like $(1-x)^3$ when $x\to 1$.

Going beyond the MV model, the gluon distribution of the Pomeron is subjected to
two types of quantum evolution: the high-energy evolution with $Y_{\mathbb{P}}$, as described by the BK/JIMWLK equations
\cite{Balitsky:1995ub,JalilianMarian:1997jx,JalilianMarian:1997gr,Kovner:2000pt,Iancu:2000hn,Iancu:2001ad,Ferreiro:2001qy,Kovchegov:1999yj} (here, applied to the gluon dipole amplitude $\mcal{T}_{g}(R,Y_{\mathbb{P}})$),
and the DGLAP evolution of the gluon distribution $x G_{\mathbb{P}}^A(x,x_{\mathbb{P}},P_{\perp}^2)$ with increasing $P_\perp^2$  \cite{Gribov:1972ri,Altarelli:1977zs,Dokshitzer:1977sg}. 

The high-energy evolution introduce a non-trivial dependence upon $Y_{\mathbb{P}}$ (or $x_{\mathbb{P}}$), but does not alter
the general structure in the r.h.s. of \eqref{xgp} --- its mere effect is to increase the saturation momentum 
and to slightly change the form of the function $\kappa$.  
In this work, we shall not consider this evolution, since in our subsequent applications to the phenomenology of UPCs we will
be led to consider only moderate values\footnote{We recall that the high energy evolution is generally assumed to start
at $x_{\mathbb{P}}= 0.01$, corresponding  $Y_{\mathbb{P}}\simeq 4.5$.} $Y_{\mathbb{P}}\lesssim 5$. 

\begin{figure}
	\begin{center}
		\includegraphics[width=0.75\textwidth]{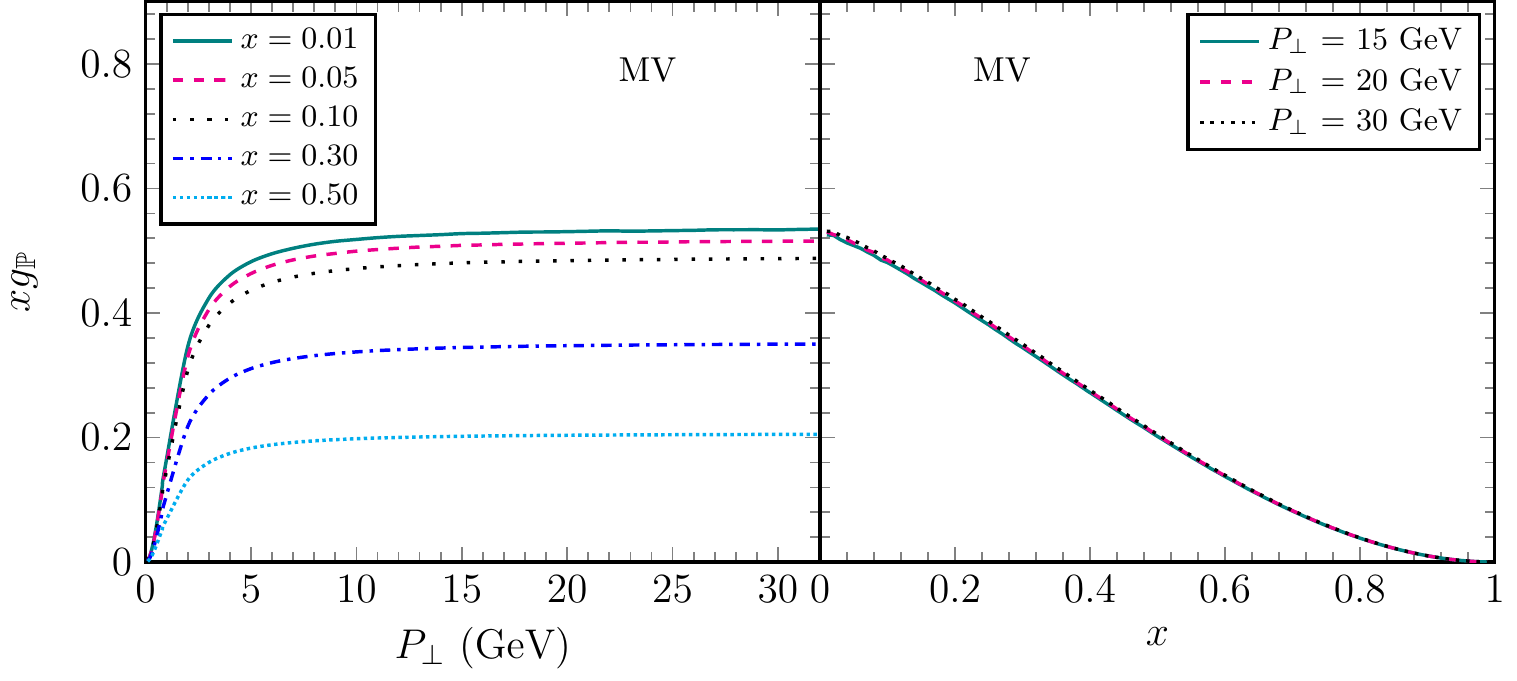}
		\includegraphics[width=0.75\textwidth]{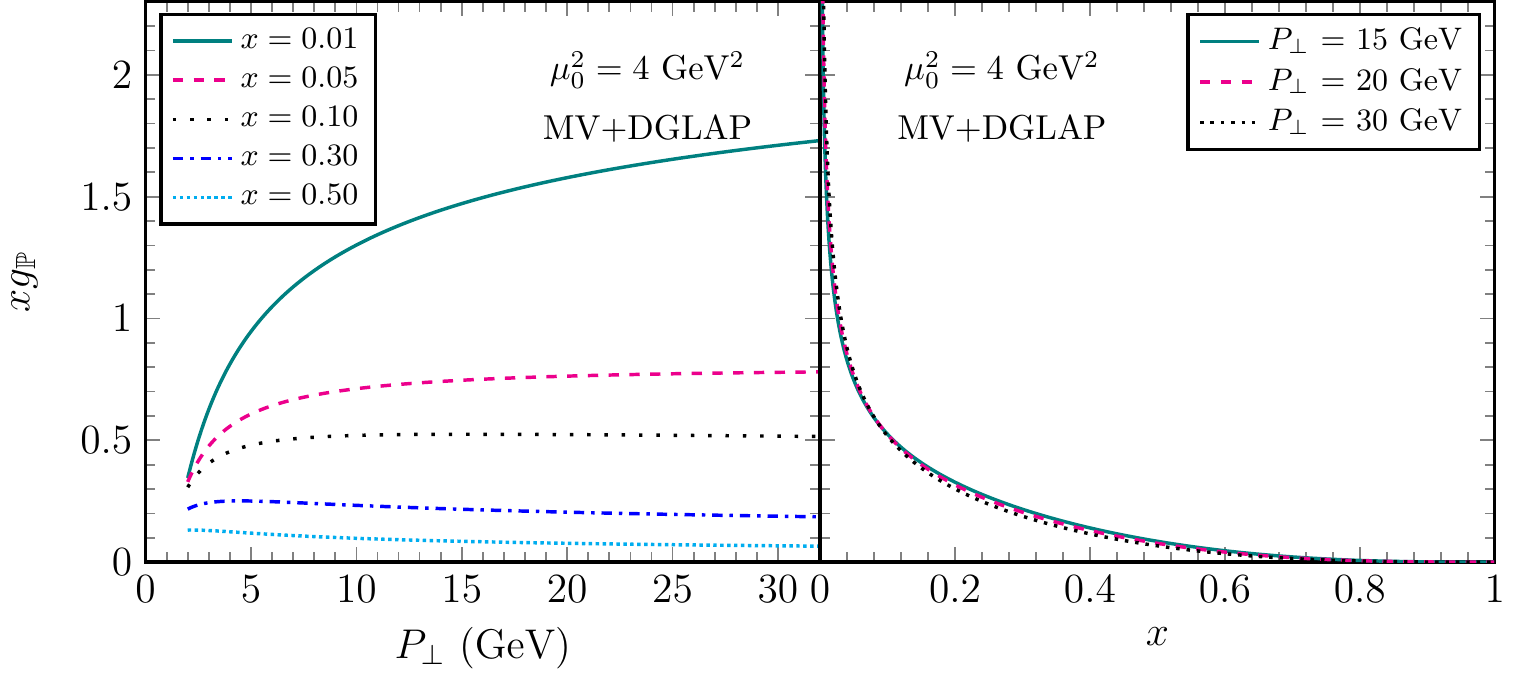}
		\includegraphics[width=0.75\textwidth]{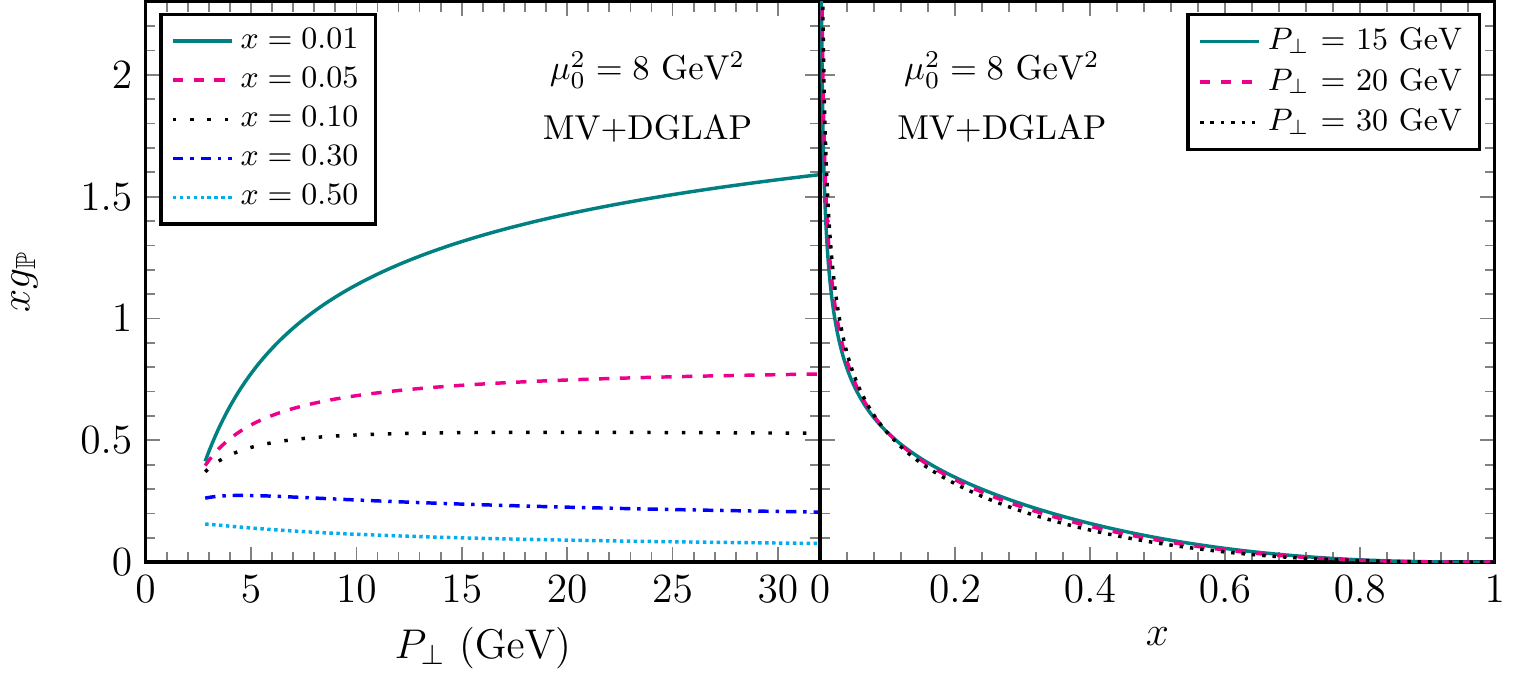}
	\end{center}
	\caption{The reduced gluon DPDF shown
	 as a function of $P_{\perp}$ for various values of $x$ and as a function of $x$ for various values of $P_{\perp}$. 
	 First line: The predictions of the MV model in  Eqs.~\eqref{MV} and \eqref{Qgs}, with $Q_{gs}^2=2$ GeV$^2$ and $\Lambda=0.2$ GeV.
	 Second and third lines: Adding the effects of the DGLAP evolution, initiated at the scale $\mu_0^2=4$ GeV$^2$ and $\mu_0^2=8$ GeV$^2$ respectively.}
\label{fig:xgp}
\end{figure}

On the other hand, the DGLAP evolution turns out to be important for
our purposes, since we shall consider relatively hard dijets, with $P_\perp$ as large as 30~GeV, for which
 $\alpha_s \ln (P_{\perp}^2/Q_{gs}^2) \gtrsim 1$. For the problem at hand, this amounts to solving an inhomogeneous 
version of the DGLAP equation, in which the unintegrated gluon distribution provided by the MV model (cf.~Eq.~\eqref{xgp-gen}) 
enters as a source term (see \cite{Iancu:2022lcw} for details).  In order to write down this equation and exhibit our numerical
results, it is convenient to introduce a reduced gluon distribution $xg_{\mathbb{P}}^A$ by stripping some
 ``trivial'' phase-space  factors off the original distribution $x G_{\mathbb{P}}^A$:
\begin{align}
	\label{xgpsmall}
	x g_{\mathbb{P}}^A(x,x_{\mathbb{P}},P_{\perp}^2) 
	\equiv 
	\frac{x G_{\mathbb{P}}^A(x,x_{\mathbb{P}},P_{\perp}^2) }{F_0}
		\qquad \mathrm{with} \qquad
	F_0\equiv\frac{N_c^2-1}{4\pi^3}\,S_{\perp} Q_{gs}^2,
\end{align}
with $Q_{gs}$ the value of the saturation momentum at tree-level (here, as given by the MV model,
cf. \eqn{Qgs}). Since we ignore the effects of the high-energy evolution, this scale  $Q_{gs}$ 
remains the physical saturation momentum
throughout our analysis (it is not affected by the DGLAP evolution).

 The relevant version of the DGLAP equation reads as follows \cite{Iancu:2022lcw}:
\begin{align}
	\label{dglap}
	\frac{\rmd 
	xg_{\mathbb{P}}^A(x, x_{\mathbb{P}}, P_{\perp}^2)}
	{\rmd \ln P_{\perp}^2} = 
	\pi P_{\perp}^2\, 
	\frac{\big|\mcal{G}^A(x,x_{\mathbb{P}},P_{\perp})\big|^2}
	{2\pi (1-x) Q_{gs}^2}+
	\frac{\alpha_s (P_{\perp}^2)}{2\pi} 
	\int_x^1 \rmd z\,
	P_{gg} (z) \,
	\frac{x}{z}\,
	g_{\mathbb{P}}^A
	\left(\frac{x}{z}, x_{\mathbb{P}}, P_{\perp}^2
	\right),	
\end{align}
where $P_{gg}(z)$ is the gluon-gluon splitting function (see e.g. \cite{Kovchegov:2012mbw}) and $\alpha_s(P_{\perp}^2)= 1/[b_0\ln(P_{\perp}^2/{\Lambda^2})]$ with
$b_0 = (11 N_c - 2  N_f)/12 \pi$ and $\Lambda=0.2$~GeV is the one-loop QCD running coupling. The DGLAP
evolution is turned on at some initial value $P_{\perp}^2=\mu_0^2$, which must obey $\mu_0^2\gg Q_{gs}^2$ and
 $\alpha_s \ln (\mu_0^2/Q_{gs}^2) \ll 1$, but is otherwise arbitrary. So long as the final momentum $P_{\perp}^2$ that
 we are interested in is much larger, $P_{\perp}^2\gg\mu_0^2$, the scheme dependence upon $\mu_0$ is expected to be small.

Using equations \eqref{Gcal} and \eqref{dglap}, we have numerically computed the reduced gluon distribution $xg_{\mathbb{P}}^A$,
with the results shown in Fig.~\ref{fig:xgp}. The two plots in the first line illustrate its functional dependences upon $P_\perp$
and upon $x$ predicted by the MV model.  The left plot shows that the distribution rapidly saturates when increasing
$P_\perp$ above  $\tilde Q_{gs}(x)$, in agreement with the above discussion of \eqn{xgp}. In particular, the $x$--dependence of
the effective saturation momentum can be appreciated from this figure. The right plot showing the $x$--dependence confirms
the $(1-x)^2$ behaviour near $x=1$ and also shows that the function $\kappa$ introduced in  \eqn{xgp} is roughly
linear in $x$, as originally noticed in  \cite{Hatta:2022lzj}.

In the second line of Fig.~\ref{fig:xgp}, we show the effects of the DGLAP evolution for an initial scale $\mu_0^2=4$~GeV$^2$.
By comparing with the respective plots in the first line, it becomes clear that the evolution effects are  substantial.
When increasing $P_{\perp}$, one sees an increase of $x g_{\mathbb{P}}^A$ for the smallest 
value $x=0.01$ and a mild decrease for larger values $x\ge 0.1$.
This is in agreement with the fact that the  DGLAP evolution copiously produces soft gluons with $x\ll 1$, while depleting the number of
their sources at larger values of $x$. In particular, the distribution becomes singular as $x\to 0$ and vanishes faster when $x\to 1$ 
than the MV model prediction $\propto (1-x)^2$ (see the right plots in Fig.~\ref{fig:xgp}). 
The growth of the gluon distribution at small $x$ has no phenomenological consequences, since in the kinematical range of interest,
the diffractive cross-section \eqref{sigmaint} is controlled by larger values $x\gtrsim 0.1$, as we shall see.
 But the faster approach to zero in the limit $x\to 1$ will be important for what follows. In the third line of Fig.~\ref{fig:xgp} 
 we repeat our analysis using $\mu_0^2=8$~GeV$^2$ as the starting scale for DGLAP evolution; comparing with the second line we see that the scheme dependence is indeed small for any $P_\perp\ge 5$~GeV and $x\ge 0.05$.

\section{Kinematical constraints and pseudo-rapidities}
\label{sect:kin}

The phase space for the 2+1 jet production in coherent diffraction is limited by two conditions. 
The first is the kinematical constraint $x\leq 1$,
 which in turn requires that $x_{q\bar{q}}$ be very small, since $x_{\mathbb{P}}$ is assumed to be small as well. 
 The second arises from the exponential decay of the photon flux which effectively requires $x_{\gamma} \lesssim x_{\gamma}^*$, cf. the discussion after Eq.~\eqref{xmax}. To study their consequences, 
 we shall consider the case that the two hard jets have equal pseudo-rapidities, $\eta_1=\eta_2 \equiv y$. This assumption,
which is useful in that it reduces the number of independent variables, is not unrealistic: we have already explained that $\eta_1$ and $\eta_2$ should be comparable with each other for the typical events. Then the longitudinal fractions of interest and the respective constraints become
\begin{align}
	\label{xxsym}
	x_{\gamma} \simeq
	\frac{P_{\perp}}{E_N}\, e^y
	\lesssim x_{\gamma}^*
	\quad \mathrm{and} \quad
	x = \frac{x_{q\bar{q}}}{x_{\mathbb{P}}}
	\simeq
	\frac{P_{\perp}}{E_N}\, e^{Y_{\mathbb{P}}-y}
	\leq 1.
\end{align}
These two conditions can be combined to limit the range of allowed values for $y$:
\begin{align}
	\label{yrange}
	  \frac{P_{\perp}}{E_N}\, e^{Y_{\mathbb{P}}}\,\le\,e^y\,\lesssim\, \frac{E_N}{P_{\perp}}\, x_{\gamma}^*\,,
\end{align}
The pseudo-rapidity $y$ of the hard dijets cannot be neither too large not too small\footnote{The conditions in \eqref{yrange} 
also introduce an upper limit on the Pomeron rapidity, namely $Y_{\mathbb P}\lesssim \ln[({E_N}/{P_{\perp}})^2 x_{\gamma}^*]$. E.g.
with $E_{N} =2.5$ \!TeV and $x_{\gamma}^{*} \simeq 1/60$, one finds 
$Y_{\mathbb P}^{\rm max} \simeq 6$ and $5$ for $P_{\perp} =15$ and 30 \!GeV, respectively.}. 
A very forward dijet (large and positive $y$) can requires a very forward photon, which means that $x_\gamma$ must be largish and the photon flux enters the regime of its exponential fall-off. A very backward dijet 
(large but negative $y$) requires a somewhat large minus longitudinal momentum fraction $x$ transferred from the target, but this is limited by
the condition that $Y_{\mathbb{P}}$ should be large. Besides, events with small $1-x\ll 1$ are strongly suppressed by the gluon distribution
\eqref{xgp}.  It is also important to notice that the window for the allowed values of $y$ widens with decreasing $P_{\perp}$. 

Let us preview how the rapidity space is marked by the various particles and gaps, as we move from forward to backward in the coherent diffractive events of interest. The nucleus $B$ appears in the very forward direction. Behind it, there is a large gap of the order of $\ln(1/x_{\gamma})$,
 which is ``trivial'' in the sense that there is obviously no hadron production before the photon dissociation. Moving towards central rapidities, one finds
the hard dijet system, which for the interesting values of $P_\perp$ and $Y_{\mathbb{P}}$, can be either slightly forward or slightly backward, as we shall see.
The third, gluon, jet  typically lies at backwards rapidities, a few units away from the hard dijets. Next, there is a large ``non-trivial'' gap --- the genuine
diffractive gap,  of the order of $Y_{\mathbb{P}}=\ln 1/x_{\mathbb{P}}$, associated with the colourless exchange between the $q\bar{q}g$ system and the target nucleus $A$. It extends all the way to the most backward region where nucleus $A$ is found.

To be more quantitative, let us first exhibit the pseudo-rapidities of the particles present in the final state. For the purposes of the current discussion we shall express them in terms of the photon energy $\omega$, the hard momentum $P_{\perp}$, the imbalance $K_{\perp}$ and the fraction $x$, while we always have in mind the value $E_N=2.5$ TeV for the energy per nucleon. Using the nucleon mass $M_N$ as an infrared regulator, Eq.~\eqref{rapid} gives for the two nuclei
\begin{align}
	\label{etaB}
	\eta_B = \ln \frac{\sqrt{2}P^+}{M_N} = \ln \frac{2 E_N}{M_N} = -\eta_A,
\end{align}
which leads to a value $\eta_B \simeq 8.5$. The pseudo-rapidities of the hard jets can be found by inverting the first equation in \eqref{xxsym} 
and recalling that $\omega = x_{\gamma} E_N$:
\begin{align}
	\label{eta12}
	\eta_1 = \eta_2 = y = \ln \frac{\omega}{P_{\perp}}.
\end{align}
Clearly, the common pseudo-rapidity $y$ of the two hard jets can have either sign depending on the ratio $\omega/P_{\perp}$. For sufficiently large $P_{\perp}$, comparable to the critical photon energy $\omega^*\simeq 40$~GeV, the dijet is more likely to propagate at central, or even
slightly negative rapidities, that is, in the opposite hemisphere w.r.t.~to the photon. To illustrate this point, we have displayed in the right panel in Fig.~\ref{fig:flux} the photon spectrum $ \omega({\dif N}/{\dif \omega})=\dif N/\dif y$ as a function of $y$ for various values of fixed $P_{\perp}$; it is there manifest that with increasing $P_{\perp}$ the support shrinks towards negative $y$. 
But this figure also shows that it should be possible to trigger on forward jets, say with $y \gtrsim 1$, even when they are relatively hard, 
say with a $P_{\perp}$ around 30 GeV. That would require large photon energies $\omega > \omega^*$, for which the spectrum is strongly suppressed,
yet non-vanishing. Albeit rare, such large $\omega$ events are very interesting, in that they explore higher center-of-mass energies for the $\gamma A$ collision and thus offer better possibilities to probe gluon saturation. 

Concerning the third, gluon, jet, it is more convenient to indicate its pseudo-rapidity separation from the hard dijets; using Eqs.~\eqref{x}-\eqref{xpom},
one finds
  \begin{align}
	\label{etajet}
	\Delta \eta_{\rm jet} = y - \eta_3=
	\ln \frac{2(1-x)}{x} + 
	\ln \frac{P_{\perp}}{K_{\perp}}.
\end{align}
This is positive and sizeable (since $P_{\perp} \gg K_{\perp}$) for all the interesting values of $x$, that is, whenever
$x$ is not very close to one\footnote{We recall that the regime $1-x\ll 1$ is suppressed by the 
large--$x$ behaviour of the gluon distribution, cf. \eqn{xgp}.}.

One can now easily calculate the relevant pseudo-rapidity gaps and separations by taking the appropriate differences. 
The photon gap is obtained as
\begin{align}
	\label{etag}
	\Delta \eta^B_{\gamma} 
	\equiv
	\eta_B - y = \ln \frac{2 E_N}{M_N} - \ln \frac{\omega}{P_{\perp}}
	= \ln\frac{2}{x_{\gamma}} + \ln \frac{P_{\perp}}{M_N}.
\end{align}
This is the sum of two large contributions since $x_{\gamma} \lesssim x_{\gamma}^* \ll 1$ and $P_{\perp} \gg M_N$. 

For the typical events, the diffractive gap extends between the  target nucleus $A$ and the gluon jet, and is
computed as
\begin{align}
	\label{etad}
	\Delta \eta^A_{\rm gap} = 
	\eta_3 - \eta_A = 
	Y_{\mathbb{P}} + \ln \frac{1}{1-x} + \ln \frac{K_{\perp}}{M_N},
\end{align}
where we have also used the following expression for $Y_{\mathbb{P}}$, which easily follows from \eqn{xxsym}:
\begin{align}
	\label{Ygap}
	Y_{\mathbb{P}} = 
	\ln \frac{\omega}{P_{\perp}} +
	\ln \frac{E_N}{P_{\perp}} - \ln\frac{1}{ x}\,.
\end{align}
For the representative values of $x$, say $0.1 \lesssim x \lesssim 0.5$, and with $K_{\perp}$ taking semi-hard values of the order of $Q_s$, both logarithms on the r.h.s.~in Eq.~\eqref{etad} are of order one. Hence, the pseudo-rapidity gap, which can be experimentally measured, is comparable to the Pomeron rapidity which appears in the theoretical description.

\eqn{Ygap} confirms that one can increase $Y_{\mathbb{P}}$  (or decrease $x_{\mathbb{P}}$) by increasing the COM energy squared 
$s_{\gamma A}=4\omega E_N$ for the photon-nucleus collision and/or by decreasing the relative transverse momentum $P_{\perp}$
of the hard dijets. Furthermore, for a fixed kinematics of the hard process, one can enhance $Y_{\mathbb{P}}$  
by triggering on events where the variable $x$ is not too small (say, $x >0.1$), meaning events in which the
rapidity separation $\Delta \eta_{\rm jet} $ between the hard dijet and the gluon jet
is as small as possible --- close to $\ln({P_{\perp}}/{K_{\perp}})$ (cf. \eqn{etajet}).

\begin{figure}
	\begin{center}
		\includegraphics[width=0.98\textwidth]{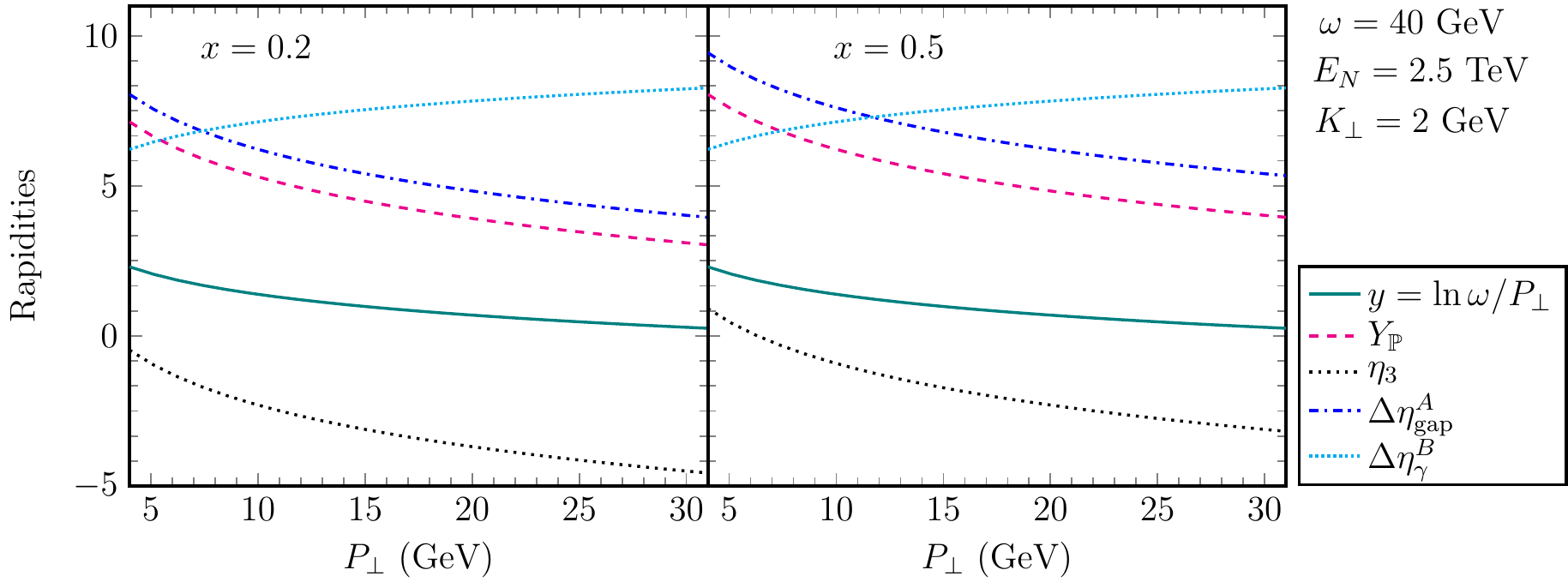}
	\end{center}
	\caption{Jet pseudo-rapidities and rapidity gaps in events with  high photon energies ($\omega =40$ GeV) as a function of the hard momentum $P_{\perp}$ for $x=0.2$ (left panel) and $x=0.5$ (right panel).}
\label{fig:gaps}
\end{figure}

To get a better feeling for all these considerations, we show the various rapidities and rapidity gaps in Fig.~\ref{fig:gaps} for some 
interesting values of the kinematical parameters. We have chosen $\omega = \omega^* \simeq 40$ GeV, a rather
large value which corresponds to somewhat rare events,
 in order to increase the probability to have small--$x_{\mathbb{P}}$ diffraction. 
 Also, we have fixed $K_{\perp}=2$ GeV, a value comparable to $Q_s$. This allows us to see how the various rapidities 
depend on the $t$-channel gluon fraction $x$ (for the 2 interesting values $x=0.2$ and $x=0.5$)
and on the hard dijet momentum $P_{\perp}$. As expected, the case for gluon saturation in the Pomeron becomes more favourable --- in the sense that the value of $Y_{\mathbb{P}}$ increases --- 
when decreasing $P_{\perp}$ and/or increasing $x$. 
Furthermore, decreasing $P_\perp$ has the effect to push the three jet system to more
forward rapidities (the more so for the third, gluon, jet), 
with the double benefit that the diffractive gap $\Delta\eta_{\rm gap}^A$ grows and the probability
to find the third jet in the rapidity range covered by the detector increases.

Let us study in more detail a couple of examples by listing explicit numbers for all the rapidities of interest. We still keep $\omega=40$ GeV and assume the more favourable value $x=0.5$ together with two relatively large values for the dijet relative momentum $P_{\perp}$ 
(15 and 30 GeV), which are experimentally accessible (in the sense of allowing for jet reconstruction) at the LHC.
\begin{enumerate}[(i)]
	\item For $P_{\perp}=15$ GeV we have
	\begin{align}
	\label{p15}
		y=1.0 \,\, | \,\, 
		\eta_3=-1.7 \,\, | \,\, 
		\Delta\eta_{\rm jet}=2.7 \,\, | \,\, 
		Y_{\mathbb{P}}=5.4 \,\, | \,\, 
		\Delta\eta_{\gamma}^B=7.5 \,\, | \,\, 
		\Delta\eta_{\rm gap}^A=6.8.
		% \,\, | \,\, \Delta\eta_{q\bar{q}}^A=9.5.
	\end{align}
	\item For $P_{\perp}=30$ GeV we have
	\begin{align}
	\label{p30}
		y=0.3 \,\, | \,\, 
		\eta_3=-3.1 \,\, | \,\, 
		\Delta\eta_{\rm jet}=3.4 \,\, | \,\, 
		Y_{\mathbb{P}}=4.0 \,\, | \,\, 
		\Delta\eta_{\gamma}^B=8.2 \,\, | \,\, 
		\Delta\eta_{\rm gap}^A=5.4.
		% \,\, | \,\, \Delta\eta_{q\bar{q}}^A=8.8.
	\end{align} 
\end{enumerate}
In Fig.~\ref{fig:jets} we depict the direction of motion of the outgoing particles in the two aforementioned cases. Both such events would be suitable for a study of gluon saturation: $Y_{\mathbb{P}}$ is large enough for that purpose (at least marginally in the second case),
although not that large to also probe the high energy evolution of the Pomeron. This motivates our study in the next section,
where we will focus on such relatively large values of $P_{\perp}$, for which we can ignore the 
BK/JIMWLK evolution with increasing $Y_{\mathbb{P}}$, but we must include the effects of the
DGLAP evolution with increasing $P_{\perp}^2$.

\section{Numerical results}
\label{sec:num}

In this section we shall present numerical results for the cross section for (2+1)-jet production via coherent
diffraction in $AA$ UPCs. 
We shall more precisely focus on the semi-inclusive cross-section \eqref{sigmaint}, in which the transverse momentum of the third jet
(which is equal to the opposite of the transverse momentum imbalance $\bK$ of the hard dijets) has been integrated out.
 As in Sect.~\ref{sect:kin}, we consider symmetric jets ($\eta_1=\eta_2 \equiv y$) 
and we rewrite  \eqn{sigmaint} in a form convenient to our purposes, namely
\begin{align}
	\label{sigmasym}
	\frac{\dif \sigma_{2+1}^{BA\to \gamma A}}
	{\dif\!\ln(1/x)\dif\eta_1 \dif \eta_2\dif^2\bP}\bigg|_{\eta_1 = \eta_2=y} =
	F_0\,
	h(P_{\perp}^2)\,
	\omega\frac{\dif N_B}{\dif \omega}\,
	x g_{\mathbb{P}}^A(x,x_{\mathbb{P}},P_{\perp}^2),
\end{align}
where we have used $\dif Y_{\mathbb{P}}=\dif\ln(1/x)$ together with \eqn{xgpsmall}.
The hard factor $h(P_{\perp}^2)$ is the limit of Eq.~\eqref{hfactor} for
 $\eta_1 = \eta_2=y$, that is, $h(P_{\perp}^2)={\alpha_{em}\alpha_s}\big(\sum e_{f}^{2}\big)/{8P_{\perp}^4}$.
Since its $P_{\perp}$-dependence is both simple and explicit, we also define a reduced, dimensionless,  cross section
by removing the factors $h(P_{\perp}^2)$ and $F_0$ from \eqn{sigmasym}
(below, the condition $\eta_1 = \eta_2=y$ is kept implicit, to simplify writing):
\begin{align}
	\label{sigmared}
	\frac{\dif \hat{\sigma}_{2+1}^{BA\to \gamma A}}
	{\dif\!\ln(1/x)\dif\eta_1 \dif \eta_2\dif^2\bP}= %\bigg|_{\rm sym} =
	\omega\frac{\dif N_B}{\dif \omega}\,
	x g_{\mathbb{P}}^A(x,x_{\mathbb{P}},P_{\perp}^2),
\end{align}
where $\omega = P_{\perp} e^y$ and $x_{\mathbb{P}} = e^{-Y_{\mathbb{P}}}$, with $Y_{\mathbb{P}}$ given by Eq.~\eqref{Ygap}. 
As anticipated, we will ignore the high-energy evolution of the Pomeron, that is, we shall compute the 
gluon DPDF $x g_{\mathbb{P}}^A(x,x_{\mathbb{P}},P_{\perp}^2)$ by using the MV model supplemented with the 
DGLAP evolution. With this approximation, the function $x g_{\mathbb{P}}^A$ does
not explicitly depend upon  $x_{\mathbb{P}}$, but only upon $x$ and $P_{\perp}^2$, as shown in  Fig.~\ref{fig:xgp}.

\begin{figure}
	\begin{center}
		\includegraphics[width=0.49\textwidth]{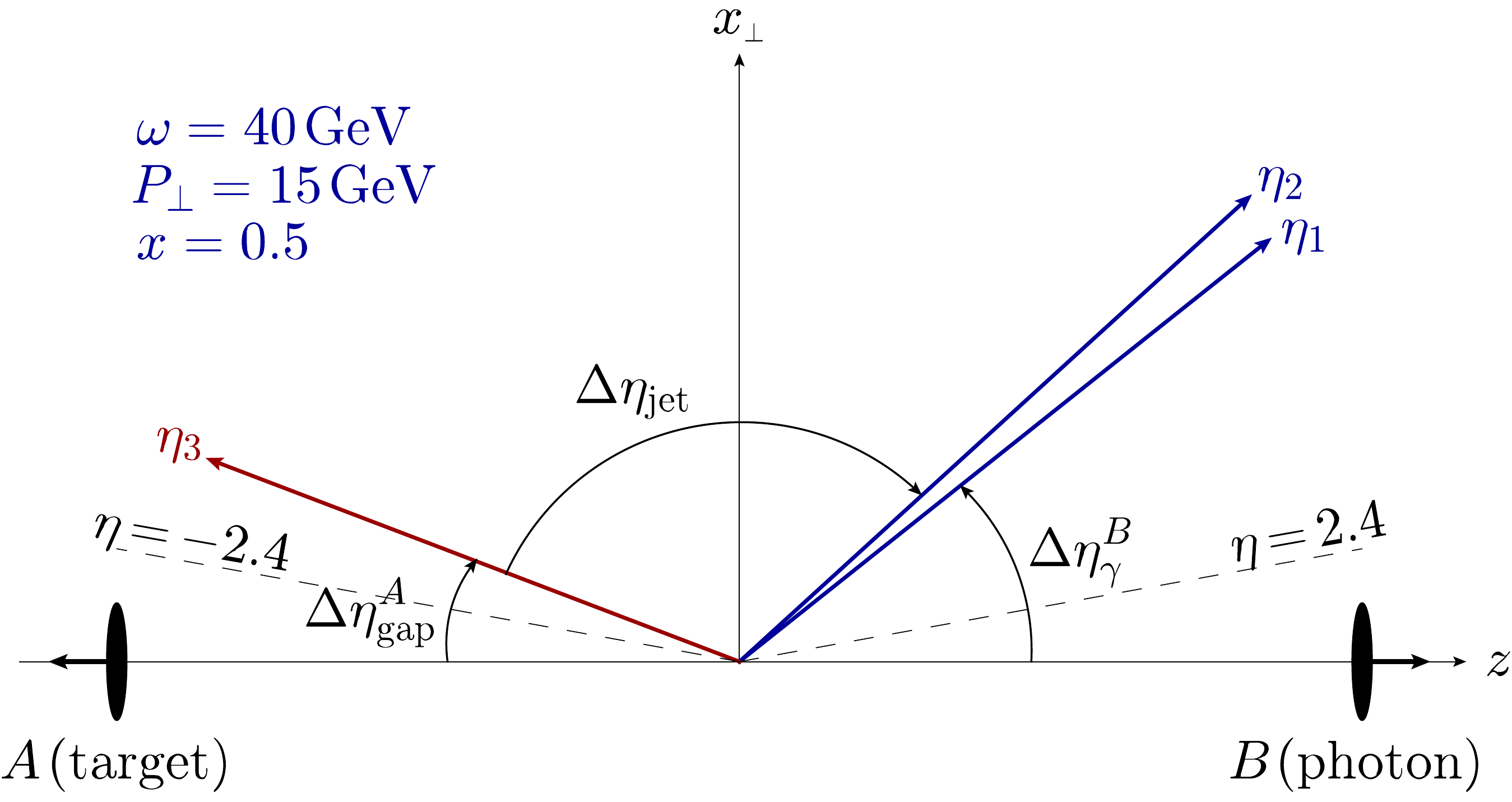}
		\hspace*{0.005\textwidth}
		\includegraphics[width=0.49\textwidth]{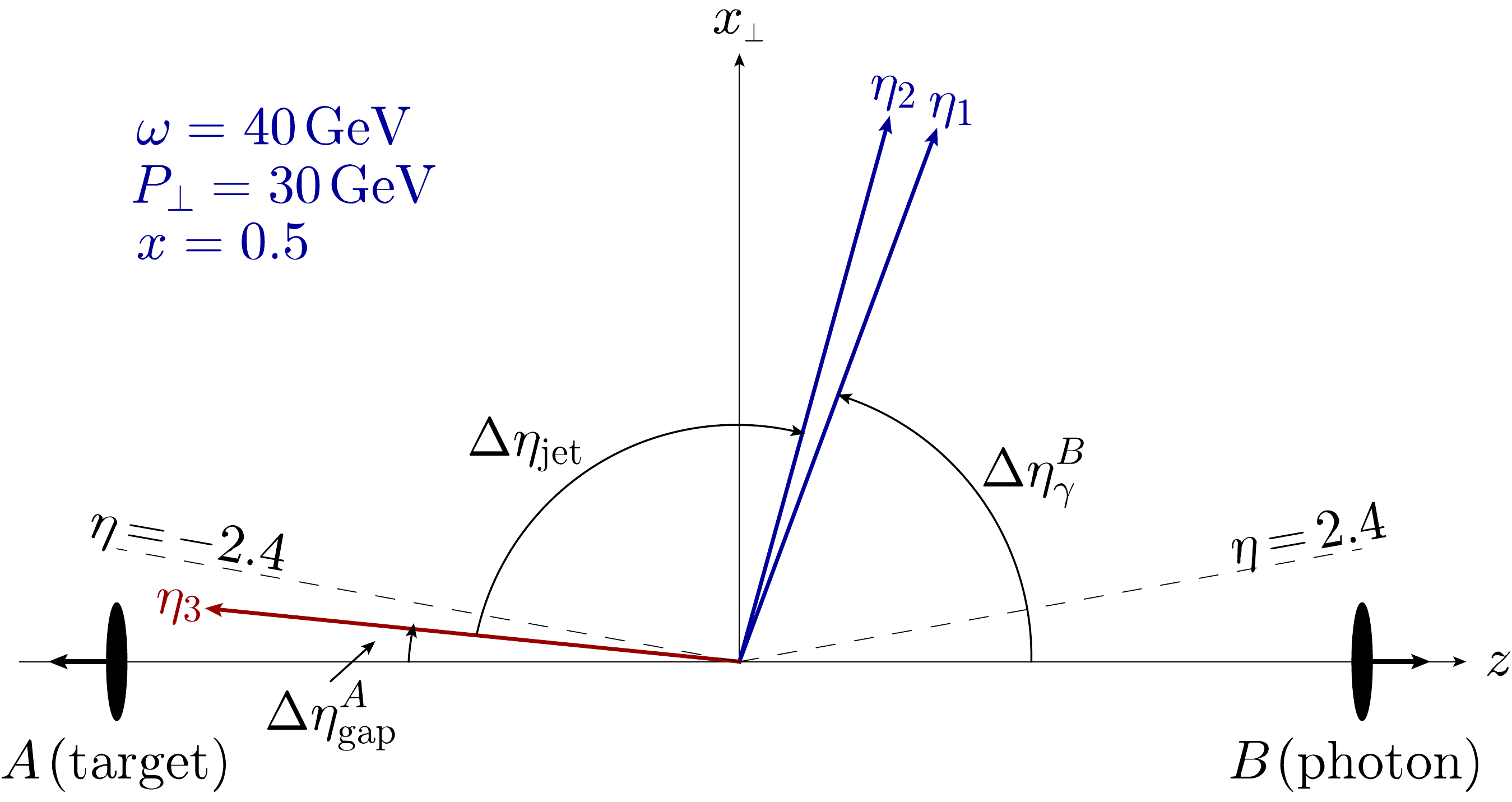}
	\end{center}
	\caption{Graphical illustration of two (2+1)--jet events which correspond to two different values of the relative momentum
	of the hard dijets: $P_{\perp}=15$~GeV (left panel) and $P_{\perp}=30$~GeV (right panel). All the other kinematical variables are
	identical and are shown in the legends. The relevant  pseudo-rapidities can be read from Eq.~\eqref{p15} (for the left panel) 
	and  Eq.~\eqref{p30} (right panel). The dotted lines in the figures represent the upper limit in pseudo-rapidity, $|\eta|=2.4$,
	of the hadronic tracker in the experimental set-up described in  \cite{CMS:2020ekd,CMS:2022lbi}.}
\label{fig:jets}
\end{figure}

Since the third jet is too soft to be reconstructed as a {\it genuine} jet in a calorimeter, the main question is whether this can be observed 
(via its hadronic descendants) by a hadronic detector. For that to be
possible, the gluon  must propagate at sufficiently central rapidities,  $|\eta_3| < \eta_0$, with $\eta_0$ the upper limit
of the pseudo-rapidity  coverage of the hadronic detector (e.g. $\eta_0 = 2.4$ at CMS  \cite{CMS:2020ekd,CMS:2022lbi}).
In our current set-up, in which the photon is a right mover and the gluon jet is almost certain to propagate in the backward hemisphere, 
the non-trivial condition reads $\eta_3 >- \eta_0$. Making use of Eq.~\eqref{etajet}, this condition translates to a lower limit 
on the gluon fraction, namely $x > x_0$ with 
\begin{align}
	\label{x0max}
	x_0 = \frac{1}{1+ \frac{K_{\perp}}{2 P_{\perp}}\,e^{y+\eta_0}}
\end{align}
This lower limit  decreases --- leading to a larger phase space --- when increasing $y$ and/or $\eta_0$ and also when decreasing $P_{\perp}$. Now we can define the ($\eta_0$-dependent) ``in'' reduced cross-section for the gluon ``jet'' to be inside the hadronic detector as
\begin{align}
	\label{sigmain}
	\frac{\dif \hat{\sigma}_{2+1, {\rm in}}^{BA\to \gamma A}}
	{\dif\eta_1 \dif \eta_2\dif^2\bP}\bigg|_{\eta_0} =
	\omega\frac{\dif N_B}{\dif \omega}\,
	\int_{x_0}^{1} \frac{\dif x}{x}\,
	x g_{\mathbb{P}}^A(x,x_{\mathbb{P}},P_{\perp}^2).
\end{align}
Strictly speaking, the lower limit $x_0$ depends on the transverse momentum $K_{\perp}$, hence the integration over $x$ must be performed prior to the one over $K_{\perp}$. In other words we should first integrate over $x$ the {\em unintegrated} gluon distribution and then integrate over $K_{\perp}$. 
This would be possible at the level of the MV model, but not also after adding the DGLAP evolution (which applies only to
the {\it integrated} distribution $x g_{\mathbb{P}}^A$).  Yet, from the discussion leading to \eqn{xgp}, we know that the integral over  $K_{\perp}$ is
controlled by momenta of the order of the target saturation momentum $Q_{gs}$ --- a scale which is not affected by the DGLAP evolution.
Hence, in practice one can evaluate the lower limit $x_0$ with a fixed value $K_{\perp}\sim Q_{gs}$ and then  \eqn{sigmain} is indeed meaningful.
We have tested this strategy in the case of the MV model and found that it gives the correct result
provided one fixes $K_{\perp} = 2$ GeV in Eq.~\eqref{x0max}.  So, we shall systematically make this choice in what follows.

\begin{figure}
	\begin{center}
		\includegraphics[width=0.45\textwidth]{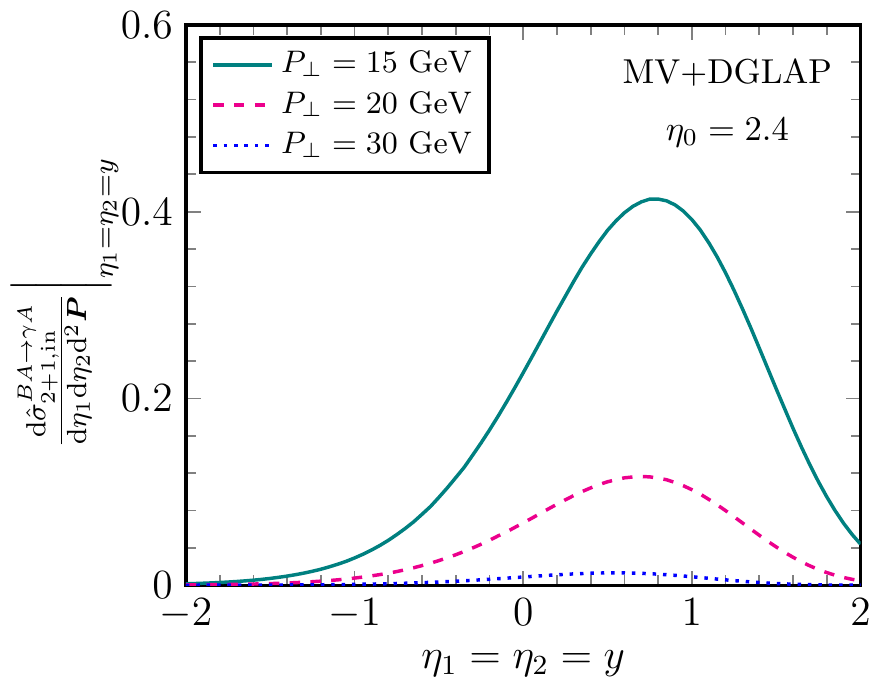}
		\hspace*{0\textwidth}
		\includegraphics[width=0.45\textwidth]{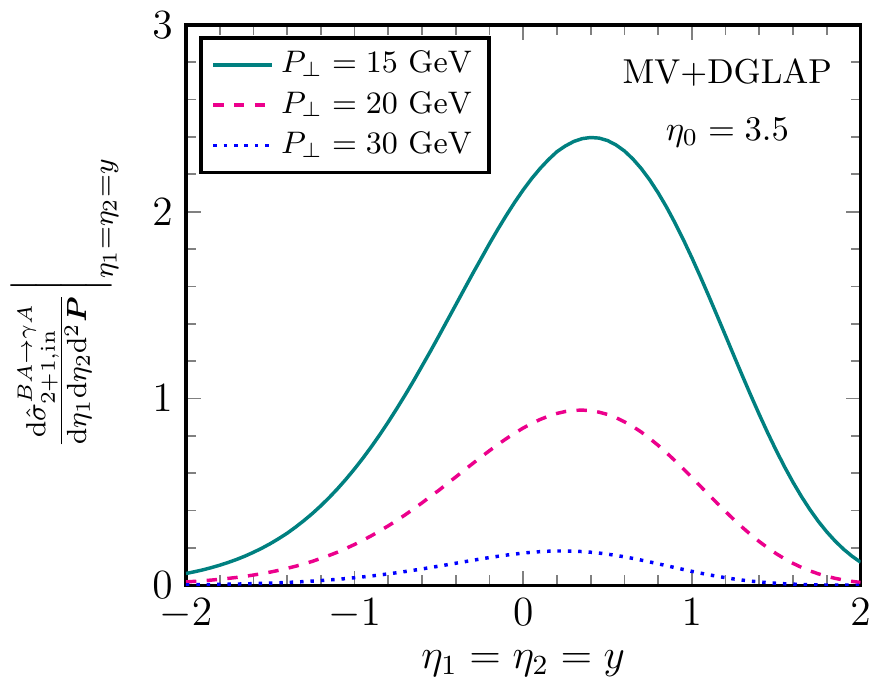}
		\\
		\includegraphics[width=0.45\textwidth]{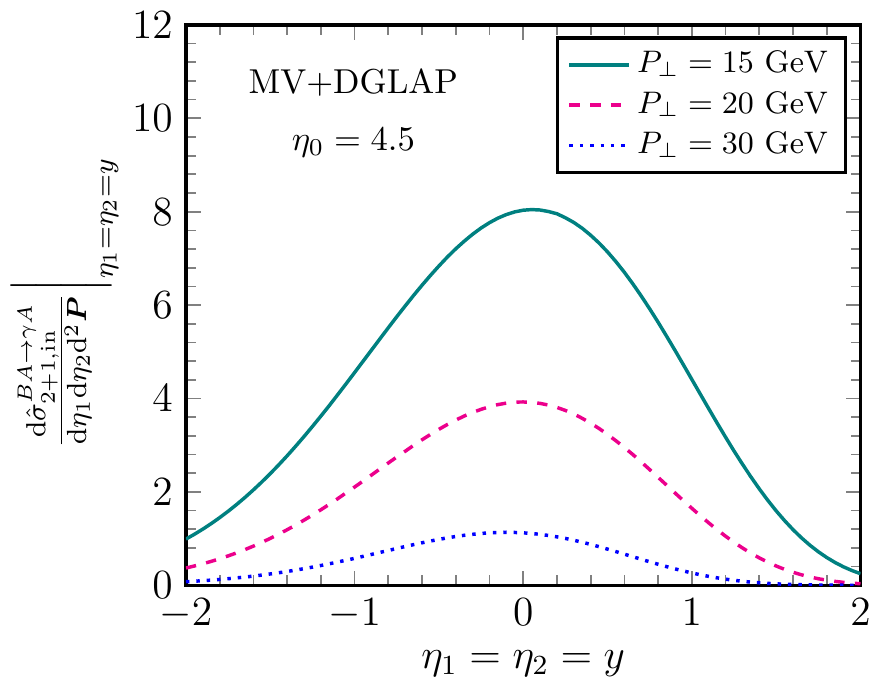}\hspace*{0\textwidth}
  \includegraphics[width=0.45\textwidth]{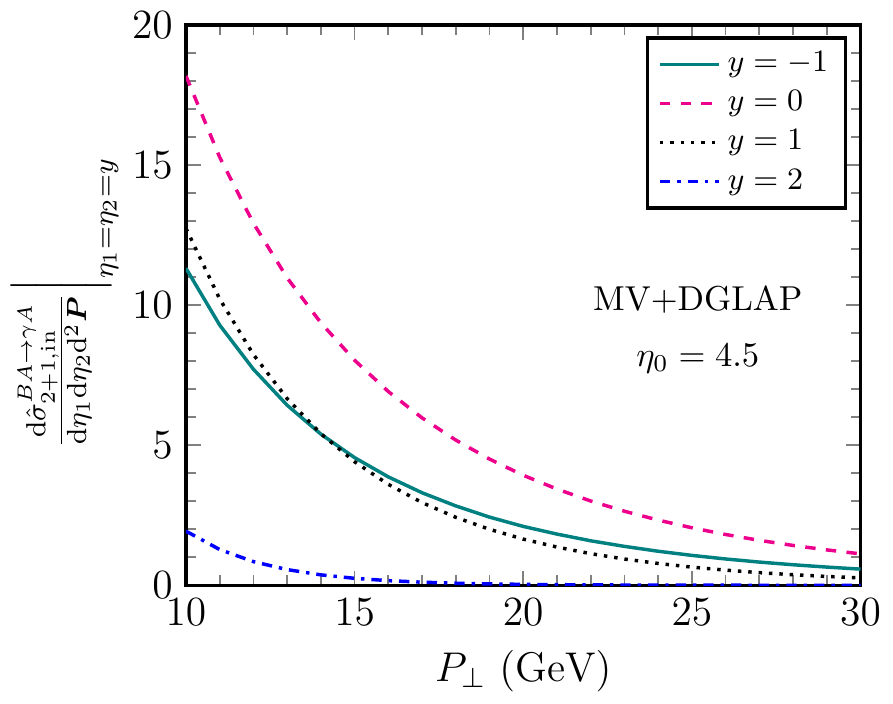}
	\end{center}
	\caption{First 3 plots: The ``in'' reduced cross-section \eqref{sigmain} 
	for finding the gluon jet in the pseudo-rapidity range $|\eta_3|<\eta_0$ 
	as a function of $y$ (the common pseudo-rapidity of the two hard jets)
	for three different values of the hard dijet momentum $P_{\perp}$ and for three values of the hadronic 
	detector's coverage: $\eta_0=2.4, \,3.5$ and 4.5.  
	(For $\eta_0 = 4.5$, the ``in'' cross-section  is by convention the same as the ``total'' cross-section.)
	The last plot: the $P_\perp$--dependence of the reduced cross-section for the particular case $\eta_0=4.5$. 
			}
\label{fig:in}
\end{figure}

The ``in'' reduced cross-section is shown in Fig.~\ref{fig:in} as a function of $y$ for $P_{\perp}=15, \,20$ and 30~GeV,
 and for three values of the detector's coverage: $\eta_0=2.4, \,3.5$ and 4.5. 
 The cross-section is suppressed at large positive $y$ due to the fast decrease of the photon flux. 
 The same is true at large negative $y$, not only because the lower limit of integration $x_0$ approaches the upper limit $x= 1$
 (so that the support of integration is squeezing), but also because the gluon distribution $x g_{\mathbb{P}}^A$ 
vanishes when $x\to 1$.  As a consequence, the cross section peaks at relatively central rapidities. 
 Furthermore, the ``in'' reduced cross-section is rapidly decreasing with $P_{\perp}$, due to the  increase in $x_0$ and also due to the 
photon flux: when increasing $P_{\perp}$ at fixed $y$, one probes  larger values for the photon energy $\omega$, where the spectrum 
 $\omega \dif N_B/\dif \omega$ is rapidly decreasing, cf. Fig.~\ref{fig:flux}. 
 To further illustrate this point, we show in the fourth plot in Fig.~\ref{fig:in} 
 the $P_\perp$--dependence of the (reduced) cross-section for the particular case $\eta_0=4.5$. 
In the actual cross-section  \eqref{sigmasym}, the suppression with increasing $P_{\perp}$ will be even faster,
due to the additional hard factor $h(P_{\perp}^2)\propto 1/P_\perp^4$.	

The largest value for $\eta_0$ considered in Fig.~\ref{fig:in}, namely $\eta_0=4.5$, 
will play a special role in what follows (and will be denoted
as $ \eta_{\scriptscriptstyle{M}}$ for more clarity): we shall use it
to conventionally define the {\it total} cross-section for producing a pair of hard dijets via the coherent (2+1)-jet channel.
The existence of an upper limit on $|\eta_3|$, or, equivalently, of a lower limit on the diffractive gap $\Delta\eta_{\rm gap}^A = |\eta_A|-|\eta_3|$, is inherent in the definition of a coherent process: such a 
process cannot exist if the diffractive gap becomes too small. Indeed, for the target nucleus $A$ not to break during the collision, it must
lose only a tiny fraction $x_{\mathbb{P}}\ll 1$ of its longitudinal momentum. 
A physically motivated condition, that was implicitly assumed in our approach, is  $x_{\mathbb{P}}\lesssim  0.02$, 
 corresponding to $Y_{\mathbb{P}}\gtrsim 4$.   In practice, we prefer to implement this
constraint on the pseudo-rapidity gap\footnote{As explained after \eqn{etad}, the difference between $\Delta\eta_{\rm gap}^A$ and $Y_{\mathbb{P}}$ 
is typically of order one.}  $\Delta\eta_{\rm gap}^A$, which can be directly measured,  unlike $Y_{\mathbb{P}}$. 
The condition $\Delta\eta_{\rm gap}^A \gtrsim 4$ together with $|\eta_A| \simeq 8.5$ imply  $\eta_{\scriptscriptstyle{M}} \simeq 4.5$, as anticipated. 
Hence,  \eqn{sigmain} with $\eta_0\to \eta_{\scriptscriptstyle{M}}= 4.5$ will be our definition for the ``total'' cross-section.
Of course, there is some ambiguity in the value of $\eta_{\scriptscriptstyle{M}}$ and varying this value (say, between 4 and 5) could be seen as a form
of ``scheme dependence''. That said, this  ambiguity is not very important for what follows, since we shall not treat this ``total'' cross-section
as a real observable\footnote{Of course, our ``total'' cross-section would become a true observable in an experiment where the
actual rapidity coverage of the hadron detector is at $|\eta|\le 4.5$.}, but only as a benchmark for estimating the relative importance of various processes.

From Fig.~\ref{fig:in}, we notice that the ``total'' cross-section 
is peaked at mid-rapidity and is rather symmetric around $y=0$. Hence, by measuring the hard dijets {\it alone}, 
it should be difficult to decide whether the photon was a right-mover, or a left-mover. 
The situation becomes considerably clearer if the third jet is also measured (most likely via its hadronic descendants). If the third jet
is more backward than the hard dijets, i.e. if the difference $\Delta \eta_{\rm jet} = y-\eta_3$ is positive, then one can safely conclude 
that the photon was a right-mover, like in our current set-up. Vice-versa, negative values for 
$\Delta \eta_{\rm jet}$ should indicate a situation where the photon was a left-mover.
So, we are mainly interested in situations where the third jet propagates at sufficiently central rapidities to be
captured by the detector ($|\eta_3|< \eta_0$). This is also the most favorable case for a study of gluon saturation,
since it corresponds to a large rapidity gap $\Delta\eta_{\rm gap}^A = |\eta_A|-|\eta_3|$.
 Moreover, it is preferable to observe this jet somewhere 
in the middle of the detector, rather than towards its edge: indeed, besides the third jet, 
we would like to also see the beginning of the gap 
at larger rapidities (within the interval $|\eta_3| < |\eta| < \eta_0$).

\begin{figure}
	\begin{center}
		\includegraphics[width=0.45\textwidth]{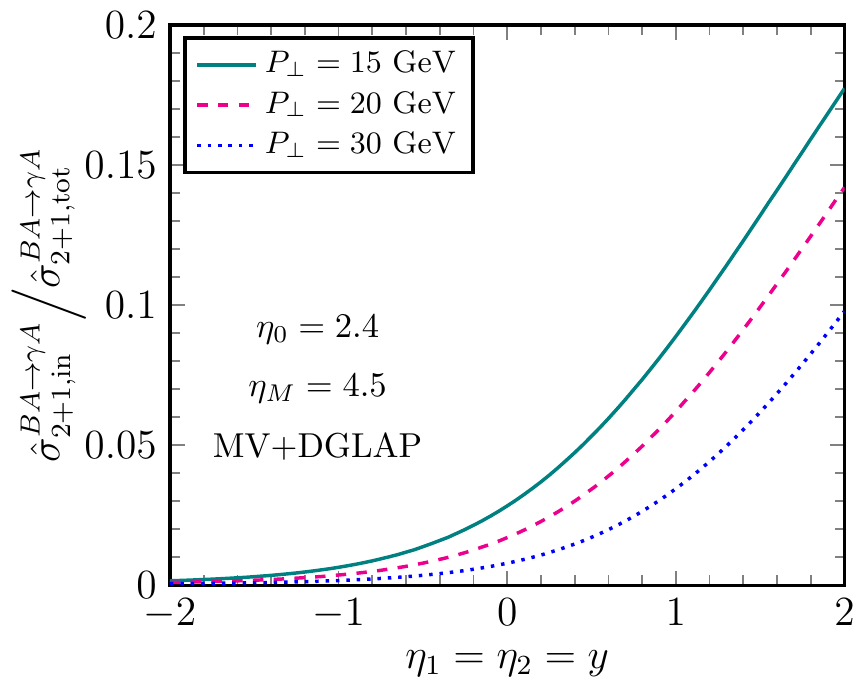}
		\hspace*{0\textwidth}
		\includegraphics[width=0.45\textwidth]{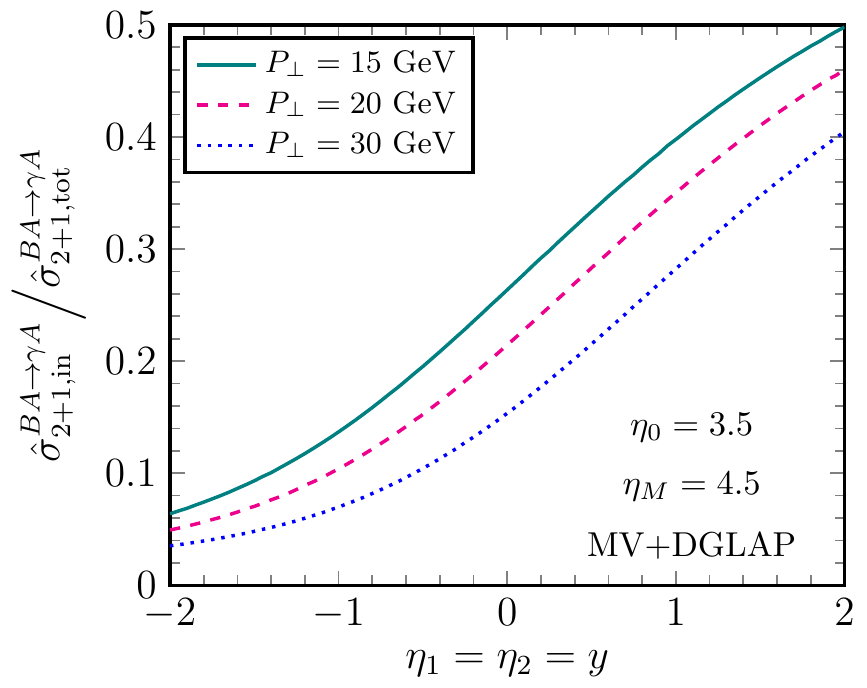}
	\end{center}
	\caption{Left figure: The ratio between the ``in'' cross-section corresponding to $\eta_0=2.4$ 
	 and the ``total'' cross-section. Right figure:  the same as in the left figure but for $\eta_0=3.5$.
		}
\label{fig:ratio}
\end{figure}

The probability for such interesting events  can be estimated as the ratio between 
the ``in'' cross-section \eqref{sigmain} corresponding to $\eta_0< \eta_{\scriptscriptstyle{M}}=4.5$
and the  ``total'' cross-section  ($\eta_0\to \eta_{\scriptscriptstyle{M}}$). This is shown in Fig.~\ref{fig:ratio} 
 for $\eta_0=2.4$ (left panel) and for $\eta_0=3.5$ (right panel).
The photon flux cancels in this ratio, which therefore is not suppressed when $y \gtrsim 1$. 
For $\eta_0=2.4$, the ratio is quite small, which may explain the difficulty to observe this jet 
in a recent measurement by CMS \cite{CMS:2020ekd,CMS:2022lbi} (see also Fig.~\ref{fig:jets}). 
That said, the ratio is significantly rising when increasing either 
the pseudo-rapidity $y$ of the hard dijets, or the detector acceptance $\eta_0$. Hence, one can enhance the 
chances to observe the third jet by triggering on events where the hard dijets are as forward as possible.

\begin{figure}
	\begin{center}
		\includegraphics[width=0.46\textwidth]{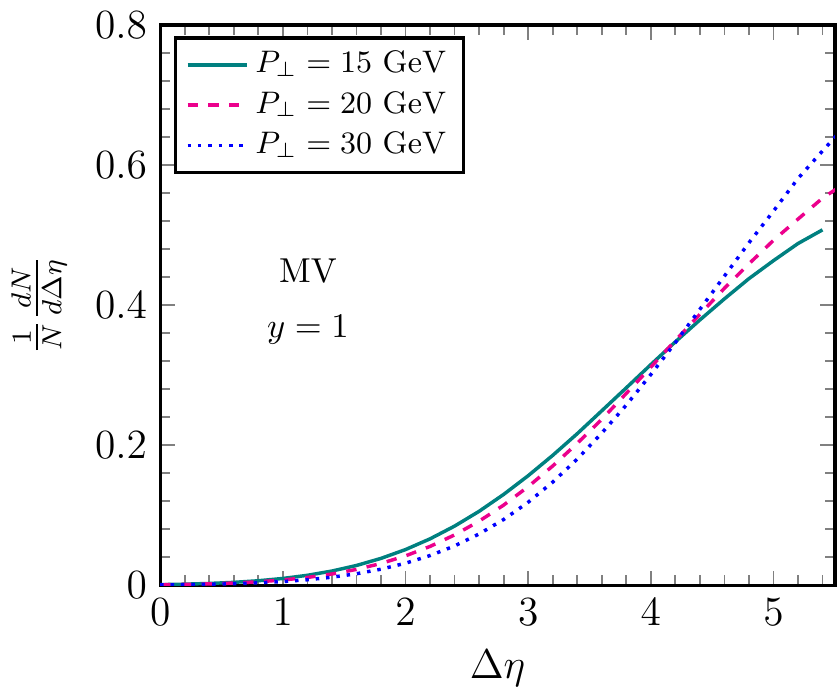}
		\hspace*{0.05\textwidth}
		\includegraphics[width=0.46\textwidth]{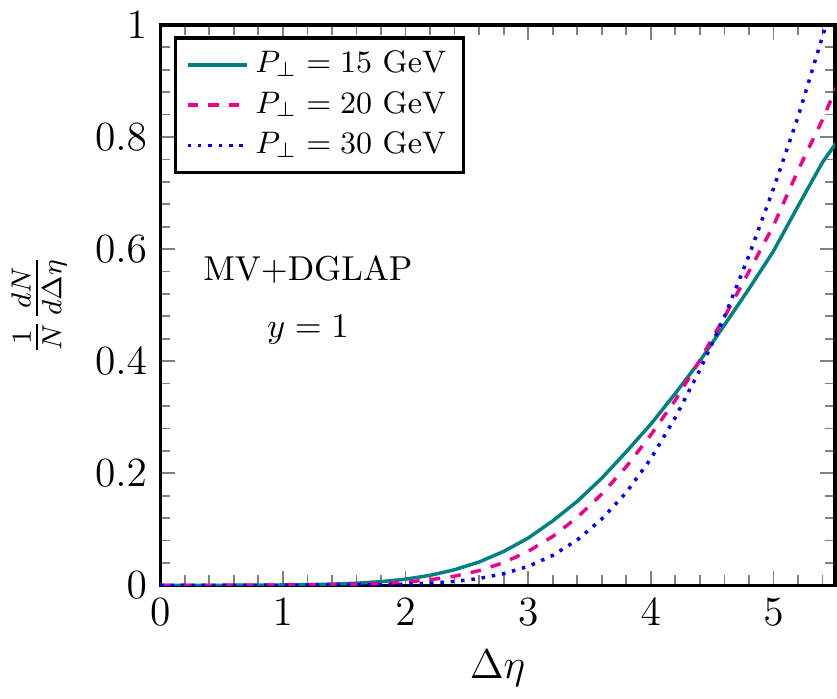}
		\end{center}
	\caption{The distribution of the pseudo-rapidity difference $\Delta \eta_{\rm jet}$  between
the gluon jet and the hard dijets, for the particular case where the hard dijets propagate at $\eta_1=\eta_2=1$.  
Left: MV model. Right: adding DGLAP evolution.}
\label{fig:eta3}
\end{figure}

For instance, for  $P_\perp=15$~GeV and $\eta_0= 2.4$, the optimal value of $y$ is $y\simeq 1$, since this yields the largest
cross-section according to the  respective plot in Fig.~\ref{fig:in}. Motivated by this observation, we show in Fig.~\ref{fig:eta3}
the distribution of the third jet in $\Delta \eta_{\rm jet} = y - \eta_3$ (the pseudo-rapidity difference between
the gluon jet and the hard dijets, cf. \eqn{etajet}) for the particular case $y=1$. This distribution  is defined as
\begin{align}
	\label{sigmadeltaeta}
	\frac{1}{N}\,\frac{\rmd N}{\rmd \Delta \eta_{\rm jet}}	\equiv
	 \frac{\dif \sigma_{\rm 2+1}^{BA \to \gamma A}
	 /\dif \eta_1 \dif \eta_2 \dif \eta_3\dif^2 \bP}
	 {\dif \sigma_{2+1,{\rm tot}}^{BA\to \gamma A}
	 /\dif\eta_1 \dif \eta_2\dif^2\bP}\,=\,\frac{(1-x)\,x g_{\mathbb{P}}^A(x,x_{\mathbb{P}},P_{\perp}^2)}
	  {\int_{x_0}^{1} \frac{\dif x'}{x'}\, x g_{\mathbb{P}}^A(x',x_{\mathbb{P}},P_{\perp}^2)},
		\end{align}
where it is understood that $\eta_1=\eta_2 \equiv y$ and $\eta_3= y - \Delta \eta_{\rm jet} $, with $y=1$.
As before, the ``total'' cross-section in the denominator is given by \eqn{sigmain} with $\eta_0=4.5$.
The value of $x$ in the numerator is related to $\Delta \eta_{\rm jet} $ via 
\beq \label{xeta3}
x =\frac{1}{1 + \frac{K_{\perp}}{2P_\perp}e^{\Delta \eta_{\rm jet}}}\,,
\eeq
(we choose $K_\perp=2$~GeV, once again) and the factor
$1-x$  comes from the Jacobian for changing rapidity variables from $\ln(1/x)$ to $\eta_3$.

The plots in  Fig.~\ref{fig:eta3} show several interesting features: \texttt{(i)} this distribution is
only weakly dependent upon $P_\perp$; \texttt{(ii)} there is a minimal value for the rapidity difference, of the order
of $\Delta \eta_{\rm jet} \simeq\ln\frac{2P_\perp}{K_\perp}$ (about 2 to 3 units of rapidity); this reflects the physics
of saturation (which fixes $K_\perp$ to be of the order of $Q_{gs}$) together with
the strong suppression of the gluon distribution near $x=1$ (an effect that is enhanced 
by the DGLAP evolution, cf. Fig.~\ref{fig:xgp}); \texttt{(iii)} the distribution in $\Delta \eta_{\rm jet}$ is rapidly growing
at larger rapidity separations $\Delta \eta_{\rm jet} \gg\ln\frac{2P_\perp}{K_\perp}$, due to the rise of 
$x g_{\mathbb{P}}$ at small $x$,  as also visible in  Fig.~\ref{fig:xgp}.

In general, this interval $\Delta \eta_{\rm jet}$ does not 
correspond to a genuine rapidity gap, since there can be hadronic activity between the hard jets 
and the third jet, notably that associated with DGLAP evolution. Yet, the DGLAP jets should be easy to distinguish from
the (original) gluon jet because, first, they are considerably harder (their transverse momenta $p_\perp$
satisfy $Q_{gs}^2\ll p_\perp^2 \ll P_\perp^2$) and, second, they are uniformly distributed in rapidity 
(e.g., if there is a single DGLAP jet,
this should be roughly located in the middle of the rapidity interval between the hard dijets and the semi-hard gluon jet).
This discussion shows that, even in situations where the third jet was not observed (since too soft or propagating at a
very large rapidity $|\eta_3|>\eta_0$), there should be still the possibility to measure the DGLAP jets and thus distinguish 
the target nucleus from the photon emitter: in pseudo-rapidity space, the DGLAP jets lie between the hard dijets and the target.

\section{Conclusions}

In this paper, we have studied semi-inclusive dijet photo-production via coherent diffraction in $AA$ UPCs and in the kinematical conditions
at the LHC. We have focused on the (2+1)--jet events  --- a hard quark-antiquark dijet and a semi-hard gluon jet 
--- which represent the dominant leading-twist contribution to diffractive dijet production in perturbative QCD. The emission of the
semi-hard gluon opens up the colour space in transverse directions and thus allows for strong scattering in the black disk limit, 
which is the necessary condition for the existence of a leading-twist contribution to diffraction. 

Due to gluon saturation, the 
transverse momentum phase-space accessible to strong scattering extends up to the semi-hard scale $Q_{gs}\sim 1\div 2$~GeV 
--- the saturation momentum of the nuclear target --- rather than being confined to the 
soft, non-perturbative, sector at $K_\perp \sim \Lambda\sim 0.2$~GeV. Diffraction lives at the upper end of this phase-space,
at $K_\perp\sim Q_{gs}$.  Since $Q_{gs}$ is semi-hard, it is legitimate to study diffraction within perturbation theory in QCD. 

Our theoretical framework was the diffractive TMD factorisation emerging from the colour dipole picture and the CGC
effective theory at high energy.  Within this framework,
we have computed the cross-section for diffractive (2+1)--jet production in $AA$ UPCs
to leading order in perturbative QCD and in the presence of multiple scattering (i.e. of gluon saturation).
Gluon saturation controls the overall strength of this cross-section and its functional dependencies 
upon the various kinematical variables and the nuclear mass number $A$, so it has observables consequences for the final state.
 
Perhaps the most striking prediction refers to the distribution of the three jets in pseudo-rapidity. For the current kinematics 
in Pb+Pb UPCs at the LHC, meaning for a center-of-mass energy  $\sqrt{\snn} = 5.2$ TeV per nucleon pair and for 
hard dijets with large transverse momenta $P_\perp\ge 20$~GeV, 
the two hard jets are predicted to propagate at nearly central rapidities, while the third,
semi-hard, jet should be separated from them by a rapidity interval 
$\Delta \eta_{\rm jet} \gtrsim \ln({2P_\perp}/{Q_{gs}})\sim 2\div 3$. 

Because of this large rapidity separation, 
and also of its relatively small transverse momentum $K_\perp\sim  Q_{gs}$, the third jet  is a priori difficult to measure
--- indeed, it has not been reported by the recent dedicated analyses at the LHC \cite{ATLAS:2017kwa,ATLAS:2022cbd,CMS:2020ekd,CMS:2022lbi}. 
Yet, its observation (e.g. as a leading hadron) would be extremely useful for the interpretation of the events
and also for comparing with the theory predictions. For instance, it would enable us to distinguish the nucleus which
has emitted the photon from that which has interacted with it. Our calculations show that the third jet propagates opposite
to the photon, in the rapidity interval between the hard dijets and the nuclear target.

Both the observability of the third jet and
the experimental study of gluon saturation would be greatly improved by lowering $P_\perp$, say down to values in the ballpark of
5 to 10 GeV. That would make it easier to produce very forward dijets and would also decrease their rapidity separation from
the semi-hard jet. In turn, this would substantially increase the (diffractive) rapidity gap between the third jet and the target nucleus
and thus allow for fully fledged studies of gluon saturation, including its high-energy, B-JIMWLK, evolution. Last but not least,
reducing $P_\perp$ would also diminish the effects of the DGLAP evolution between the third jet and the hard dijets, 
thus reducing the risk for confounding the third jet with one of the DGLAP jets in the experiments.

\section*{Acknowledgements} We would like to thank our colleagues experimentalists Alexandr Bylinkin, Zvi Citron, Yftach Moyal, Christophe Royon and Daniel Tapia Takaki, for useful discussions and clarifications on the experimental set-up in UPCs at the LHC. 
 The work of A.H.M. is supported in part by the U.S. Department of Energy Grant \# DE-FG02-92ER40699. 
S.Y.~Wei is supported by the Taishan fellowship of Shandong Province for junior scientists and the Shandong Province Natural Science Foundation under grant No.~2023HWYQ-011.

\appendix                                                                                                                     
 
\section{Comparing diffractive (2+1)-jets with exclusive dijets}
\label{sec:app}

We have mentioned in the Introduction that the cross section for exclusive dijet production is a higher-twist effect,
which is (roughly) suppressed by a factor $Q_{gs}^2/P_\perp^2$ compared to the cross-section for the (2+1)--jet coherent channel.
Since experimentally it seems difficult to distinguish exclusive from (2+1)--production, it becomes importany to compare the 
respective cross-sections in more detail. The exclusive dijet cross section in UPCs can be trivially obtained from the corresponding one in $\gamma^*\!A$ collisions (see for example Appendix A in \cite{Iancu:2022lcw}) by taking the real-photon limit $Q^2\to 0$ 
and  multiplying the result with the photon flux; one finds
\begin{align}
	\label{sigma2jet}
	\frac{\dif \sigma_{\rm exc}^{BA \to \gamma A}}
	{\dif \eta_1 \dif \eta_2 \dif^2 \bP} = 
	\frac{S_{\perp} \alpha_{\rm em} N_c}{2\pi^2}\,
	\left(\sum e_f^2\right)
	\vartheta_1\vartheta_2 
	\big( \vartheta_1^2 + \vartheta_2^2 \big)\,
	\omega \frac{\dif N_B}{\dif \omega}\,
	\big| \mcal{A}^l_{\rm exc} (\bP, Y_{q\bar{q}}) \big|^2.
\end{align}
Here $\vartheta_2 = 1 -\vartheta_1$, $Y_{q\bar{q}}= \ln (1/x_{q\bar{q}})$, while $\omega$, $\vartheta_1$ and $x_{q\bar{q}}$ are expressed in terms of the hard momentum $P_{\perp}$ and the pseudo-rapidities $\eta_1$ and $\eta_2$ through Eqs.~\eqref{omega}, \eqref{theta1} and \eqref{xqq}.  The amplitude $\mcal{A}^i_{\rm exc}(\bP, Y_{q\bar{q}})$ is related to a particular Fourier transform which involves the amplitude for a $q\bar{q}$ dipole to scatter of the nucleus target. For our purposes it suffices to present only the high momentum tail 
within the MV model (where the amplitude becomes independent of $Y_{q\bar{q}}$):
\begin{align}
	\label{aexc}
	\mcal{A}^l_{\rm exc}(\bP) 
	\simeq
	 - i\,\frac{Q_A^2 P^l}{P_{\perp}^4}
	\quad \mathrm{for} \quad P_{\perp} \gg Q_s.
\end{align}
Notice that both $Q_A$ and $Q_s$ in the above refer to a quark-antiquark dipole, i.e.~they are proportional to $C_F$; for instance,
$Q_A^2=({C_F}/N_c)Q_{gA}^2$, where $Q_{gA}$ is the corresponding scale for a gluon dipole, as introduced in 
Eq.~\eqref{MV}. 

\eqn{aexc} implies $\big| \mcal{A}^l_{\rm exc} (\bP) \big|^2={Q_A^4}/{P_{\perp}^6}$, hence the exclusive
dijet cross-section is of higher-twist order: it is suppressed by a factor $Q_A^2/P_{\perp}^2$ with respect to the cross-sections
for both 
inclusive dijets \cite{Dominguez:2012ad} and diffractive (2+1)-jets. Also, it scales like $A^{4/3}$ with the mass number
of the nuclear target.

\begin{figure}
	\begin{center}
		\includegraphics[width=0.45\textwidth]{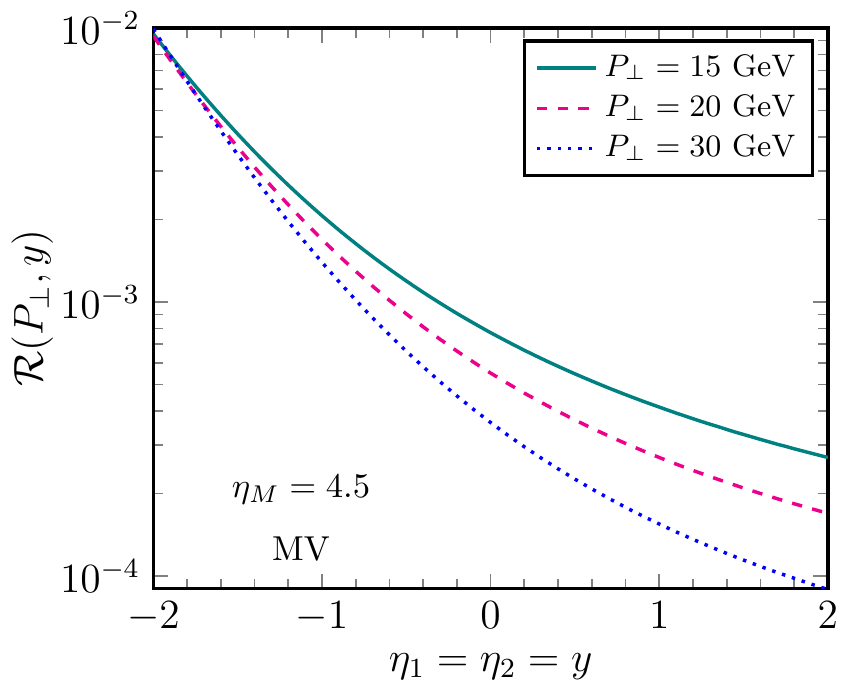}
		\hspace*{0\textwidth}
		\includegraphics[width=0.45\textwidth]{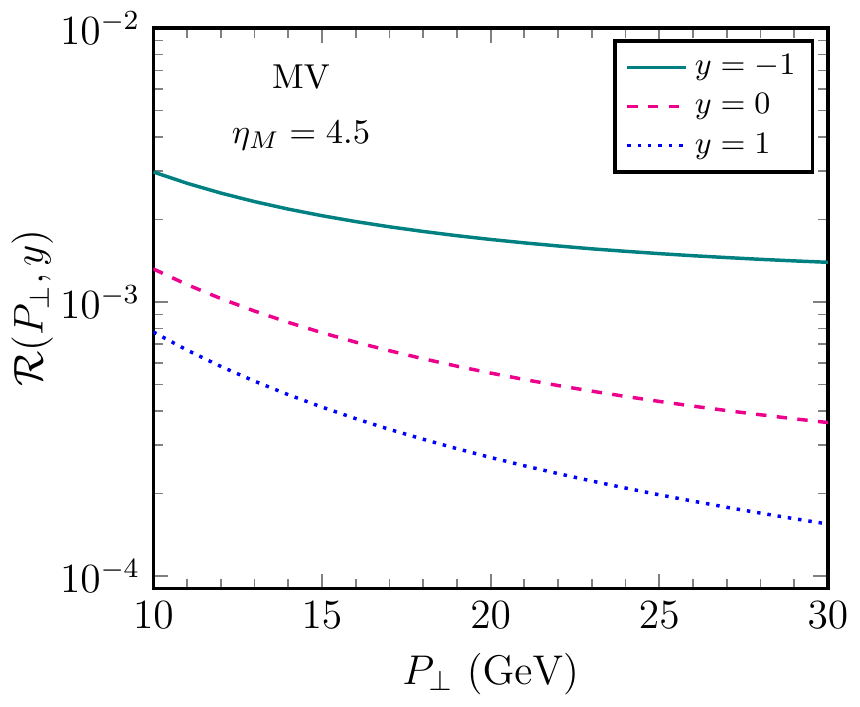}\\
		\includegraphics[width=0.45\textwidth]{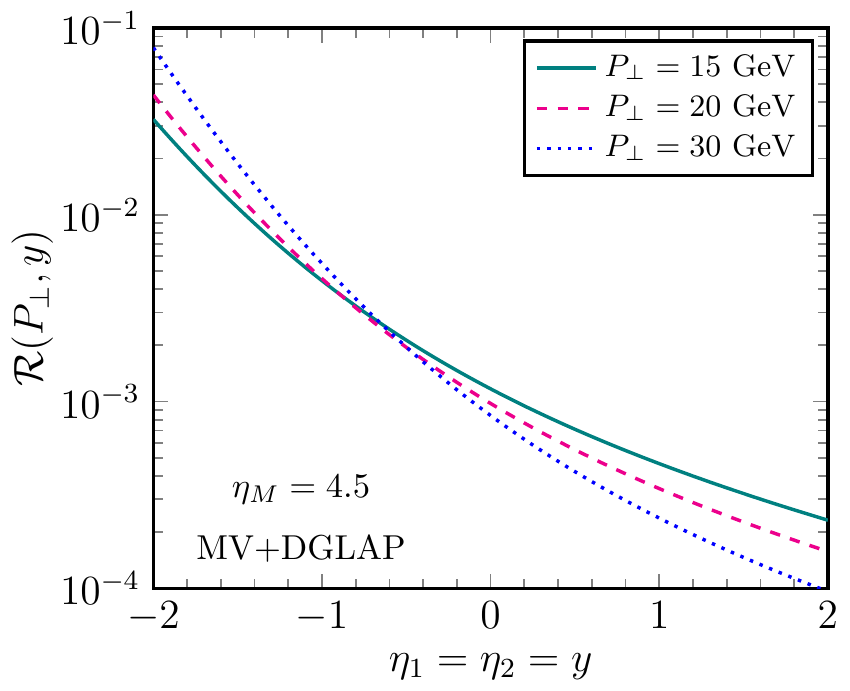}
		\hspace*{0\textwidth}
		\includegraphics[width=0.45\textwidth]{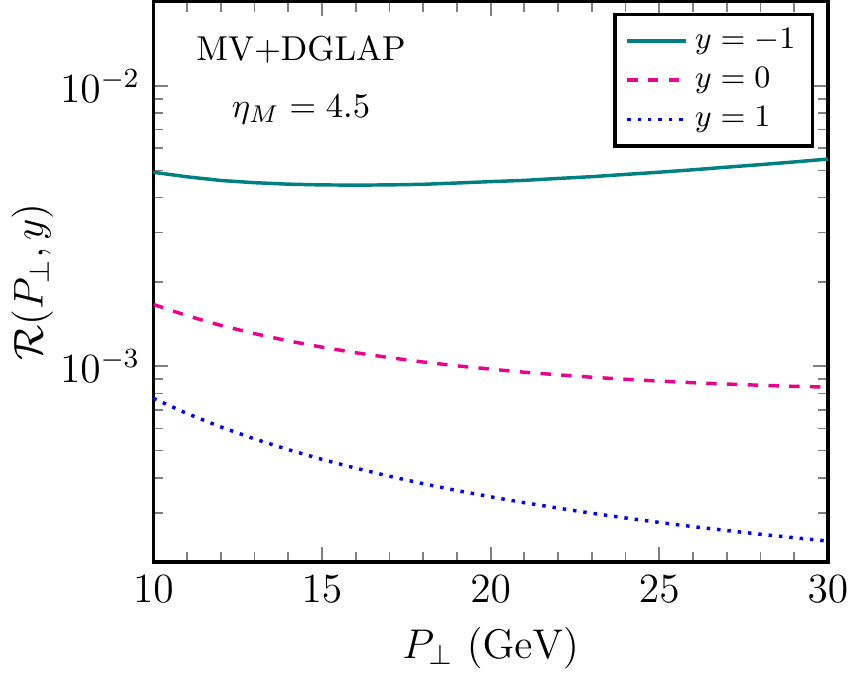}
	\end{center}
	\caption{Cross section for exclusive dijet production divided by the ``total'' cross section for 2+1 jet production as a function of $y$ for various values of $P_{\perp}$ (left panel) and as a function of $P_{\perp}$ for a various values of $y$. Upper line: MV model. Lower line: MV model +
	 DGLAP evolution. }
\label{fig:R}
\end{figure}

In order to perform the desired comparison we form the following ratio (once again,
we specialise to symmetric jets, with $\eta_1=\eta_2 \equiv y$) 
\begin{align}
\label{R}
	 \mcal{R}(P_{\perp},y)
	 \equiv
	 \frac{\dif \sigma_{\rm exc}^{BA \to \gamma A}
	 /\dif \eta_1 \dif \eta_2 \dif^2 \bP}
	 {\dif \sigma_{2+1,{\rm tot}}^{BA\to \gamma A}
	 /\dif\eta_1 \dif \eta_2\dif^2\bP}
	 = 
	 \frac{2\pi}{\alpha_s} \frac{N_c}{N_c^2-1}\,
	 \frac{P_{\perp}^4}{Q_{gs}^2}\,
	 \frac{\big|\mcal{A}^l_{\rm exc}(\bP)\big|^2}
	 {\int_{x_0}^{1} \frac{\dif x}{x}\, x g_{\mathbb{P}}^A(x,x_{\mathbb{P}},P_{\perp}^2)},
\end{align}
in which the photon flux trivially cancels. The lower limit $x_0$ in the integral over $x$ is given by \eqn{x0max} with $\eta_0= 4.5$.
The strong coupling in the above refers to the gluon emission with transverse momentum $K_{\perp} \sim Q_{gs}$ in the 2+1 jets case.
By using its one-loop version $\alpha_s(Q_{gs}^2)$, defined below Eq.~\eqref{dglap}, together with the relation  \eqref{Qgs} between $Q_{gs}$ and $Q_{gA}$
and \eqn{aexc} for $\mcal{A}^l_{\rm exc}$ in the MV model, one easily finds
\begin{align}
\label{R1}
	 \mcal{R}(P_{\perp},y)=\frac{\pi b_0}{N_c}\,\frac{C_F}{N_c}\,\frac{Q_{gA}^2}{P_{\perp}^2}\,\frac{1}{
	  \int_{x_0}^{1} \frac{\dif x}{x}\, x g_{\mathbb{P}}^A(x,P_{\perp}^2)}\,,
\end{align}
where ${\pi b_0}/{N_c}=0.75$ for $N_c=N_f=3$ and the integral in the denominator is a quantity of order one. 
Since $Q_{gA}\simeq 0.7$~GeV is much smaller than $P_\perp\ge 15$~GeV,
it is clear that the ratio is very small,  in the ballpark of $10^{-2}$ to  $10^{-3}$, depending upon the value of $P_\perp$. This is in agreement with
the numerical results presented in Fig.~\ref{fig:R}. As also visible in Fig.~\ref{fig:R}, the ratio is rapidly falling with $y$, due to the
rapid decrease of $x_0$ and to the increase of $ x g_{\mathbb{P}}^A$ at small $x$, cf. Fig.~\ref{fig:xgp}.
The would-be rapid decrease  $\propto 1/P_{\perp}^2$ introduced by the
exclusive cross-section is somewhat tempered by the fact that $x_0$ increases towards unity for very large $P_\perp$, so the integral
gets squeezed  to $x\sim 1$, where $x g_{\mathbb{P}}^A$ {\it is} suppressed. This effect becomes stronger at negative
values of $y$ (leading to larger $x_0$) and in the presence of 
the DGLAP evolution, which enhances the suppression of the gluon distribution at large $x$, cf. Fig.~\ref{fig:xgp}.
That said, the numerical results in Fig.~\ref{fig:R} confirm that  the exclusive dijet production is two to three orders of magnitude 
smaller than the 2+1 jet production in the whole kinematic regime of interest. 

%\bibliographystyle{utcaps}
%\bibliography{refs}

\providecommand{\href}[2]{#2}\begingroup\raggedright\endgroup

\end{document}